\documentclass[useAMS,usenatbib]{mn2e}
\usepackage{txfonts}
\usepackage{graphicx}
\usepackage{natbib}
\usepackage{amssymb}
\usepackage{color}

\newcommand{\chandra}{{\it Chandra}} 
\newcommand{\xmm}{{\it XMM-Newton}}
\newcommand\lax{\>\vcenter{\hbox{$<$\hskip-.75em\lower1.0ex\hbox{$\sim$}}}\>}
\newcommand\uax{\>\vcenter{\hbox{$>$\hskip-.75em\lower1.0ex\hbox{$\sim$}}}\>}

\newcommand{\sigth}{\sigma_{\rm T}}
\newcommand{\mpr}{m_{\rm p}}
\newcommand{\sigsb}{\sigma_{\rm SB}}
\newcommand{\Teff}{T_{\rm eff}}
\newcommand{\keso}{\kappa^{\rm es}_0}
\newcommand{\kfmean}{\kappa^{\rm F}}
\newcommand{\zg}{z_g}

\newcommand{\aap}{A{\&}A}

\newcommand{\apj}{ApJ}
\newcommand{\apjl}{ApJL}
\newcommand{\apjs}{ApJS}
\newcommand{\araa}{ARA{\&}A}
\newcommand{\pra}{Phys. Rev. A}

\newcommand{\pre}{Phys. Rev. E}
\newcommand{\mnras}{MNRAS}

\newcommand{\nat}{Nature}
\newcommand{\physrep}{Phys.~Rep.}

\newcommand{\azh}{AZh}

\title[Modelling mid-Z element atmospheres for strongly-magnetized
  neutron stars]{Modelling mid-Z element atmospheres for strongly-magnetized
  neutron stars}
\author[K. Mori and W.C.G. Ho]{Kaya Mori$^{1, 2}$\thanks{e-mail:
  kaya@cita.utoronto.ca}, Wynn C.G. Ho$^{3,4}$\thanks{e-mail: wynnho@slac.stanford.edu}\\
$^{1}$Department of Astronomy and Astrophysics,
   University of Toronto, 50 St. George Street, Toronto, Ontario, M5S
   3H4, Canada\\
$^2$Canadian Institute for Theoretical Astrophysics,
University of Toronto, 60 St. George Street, Toronto, Ontario, M5S 3H8, Canada\\
$^3$Harvard-Smithsonian Center for Astrophysics, 60 Garden Street,
Cambridge, MA, 02138, USA\\
$^4$Kavli Institute for Astrophysics and Space Research,
Massachusetts Institute of Technology, Cambridge, MA, 02139, USA}

\begin{document}
\pagerange{\pageref{firstpage}--\pageref{lastpage}} \pubyear{2006}

\maketitle 

\label{firstpage}

\begin{abstract}

We construct models for strongly-magnetized neutron star atmospheres
composed of mid-Z elements (carbon, oxygen and neon) with magnetic fields
$B=10^{12}$--$10^{13}$ G and effective temperatures
$\Teff=(1-5)\times10^6$ K; this is done by first addressing the physics
relevant to strongly-magnetized plasmas and calculating the equation of
state and polarization-dependent opacities.
We then obtain the atmosphere structure and spectrum
by solving the radiative transfer equations in hydrostatic and radiative
equilibrium.  In contrast to hydrogen opacities at the relevant
temperatures, mid-Z element opacities are dominated by numerous
bound-bound and bound-free transitions. Consequently, temperature
profiles are closer to grey profiles, and photosphere densities are lower
than in the hydrogen case. Mid-Z element atmosphere spectra are
significantly softer than hydrogen atmosphere spectra and show numerous
absorption lines and edges.  The atmosphere spectra
depend strongly on surface composition and magnetic field but weakly
on surface gravity. Absorption lines are primarily
broadened by motional Stark effects and the (unknown) surface magnetic field
distribution. When magnetic field variation is not
severe, substructure in broad absorption features can be resolved by 
(phase-resolved) CCD spectroscopy from {\it Chandra} and {\it XMM-Newton}.
Given the multiple absorption features seen in several isolated
neutron stars, it is possible to determine the surface composition, magnetic
field, temperature, and gravitational redshift with existing X-ray data;
we present qualitative comparisons between our model spectra and the
neutron stars 1E1207.4$-$5209 and RX~J1605.3$+$3249.
Future high-resolution X-ray missions such as {\it Constellation-X} will
measure the gravitational redshift with high accuracy by resolving narrow
absorption features; when combined with radius measurements, it will
be possible to uniquely determine the mass and radius of isolated neutron
stars. 

\end{abstract}



\section{Introduction} 
\label{sec_intro}

Recent observations by \chandra\ and \xmm\ detected spectral
features from isolated neutron stars (INSs). A single or multiple absorption
features have been found from nearby radio-quiet NSs \citep{haberl03,
vankerkwijk04, haberl04, zane05, schwope05, haberl06_2} and two broad absorption
features were detected from 1E1207.4$-$5209
\citep[hereafter 1E1207;][]{sanwal02, mereghetti02, mori05}.
In fact, 1E1207 is unique among these INSs since one of
the features appears above 1 keV, while the other INSs have absorption
features at $E\simeq0.2$--$0.7$ keV. Identification of the observed spectral
features will not only allow a measurement of the surface composition and
magnetic field strength $B$ but also constrain the nuclear equation of
state via a measurement of the gravitational redshift $\zg$, where
$\zg$ is given by $(1+\zg)=(1-2GM/Rc^2)^{-1/2}$ and $M$ and $R$ are the
NS mass and radius, respectively.

The surface composition is usually assumed to be hydrogen because the fast
sedimentation time (due to the strong surface gravity) causes the lightest
elements to rise to the top \citep{alcock80} and only a tiny amount of
hydrogen is required to produce an optically thick atmosphere
\citep{romani87}. Hydrogen atmospheres in strong magnetic fields have
been studied in great detail by addressing the complicated properties
of strongly-magnetized dense plasmas \citep{pavlov95_1, lai97, potekhin99,
potekhin03, potekhin04}.  Partially-ionized hydrogen atmosphere models
show spectral features, such as the proton cyclotron line, photo-ionization
edge, and atomic transition lines at $E\la1$~keV \citep{ho03, potekhin04_2}.
However, hydrogen atmospheres cannot produce strong spectral
features at $E\ga 1$~keV because (1) the binding energy of a hydrogen
atom never exceeds $\sim1$ keV at any $B$ \citep{lai01, sanwal02}, (2) QED
effects significantly reduce line strengths at $B\ga7\times10^{13}$~G
\citep{ho02, ho04, vanadelsberg06}, and (3) the fraction of hydrogen
molecular ions
\citep[which may have transition lines above 1 keV;][]{turbiner04_1}
is negligible ($<10^{-6}$) at $B\la10^{15}$ G and
$T\ga10^6$ K \citep{potekhin04}.  Therefore,
spectral features at $\ga 1$~keV (such as the 1.4~keV absorption line
of 1E1207) hint at a non-hydrogenic element
atmosphere on the NS surface \citep{sanwal02, hailey02, mori06}. Also, recent
theoretical studies suggest that a hydrogen layer on the
surface may be rapidly depleted by diffuse nuclear burning or pulsar
winds, thus exposing the heavier elements that lie underneath
\citep{chang04_1, chang04_2}.

The few existing non-hydrogenic atmosphere models
\citep{miller92, rajagopal97} are far from complete, due mainly to a
lack of accurate atomic data for multi-electron ions in the high
magnetic field regime and crude treatment of ionic motional
and non-ideal effects.  In this paper, we construct mid-Z
element (carbon, oxygen and neon) atmosphere models that are
compatible with the high quality X-ray data of \chandra\ and \xmm.
We only consider magnetic field strengths $B=10^{12}$--$10^{13}$~G
and effective temperatures $\Teff=(1-5)\times10^6$~K because
QED effects may become important at higher magnetic fields
\citep[and references therein]{ho04} and molecules may become abundant
at lower temperatures
\citep{medin06_1}. QED effects on mid-Z element atmosphere structure
and spectra will be studied in future work \citep{ho06_2}.  We also
defer investigation of strongly-magnetized helium atmospheres due to
non-trivial motional Stark effects and possible contamination by
helium molecules \citep{mori06_3}.

We briefly describe here several major improvements of our mid-Z
element atmosphere models over the previous atmosphere models by
\citet{miller92} (for helium and mid-Z elements) and
\citet{rajagopal97} (for iron).

(1) The previous non-hydrogenic atmosphere models assume that bound
electrons are in the ground Landau state (adiabatic
approximation). However, the adiabatic approximation is not valid for
the low $B$ and high atomic number $Z$ regime that we consider in this
paper.  For instance, the ionization energy of the innermost electron of
$Z=6$--10 element ions at $B=10^{12}$ G is underestimated by 5--10\%
in the adiabatic approximation.
     
(2) Numerous bound states and transitions are needed to construct
mid-Z element atmospheres. The model of \citet{miller92} does not
include bound-bound transitions, while the iron atmosphere model of
\citet{rajagopal97} utilizes rather crude atomic line data (the energy
values and oscillator strengths have as much as 10\% and a factor of 2
uncertainties, respectively). We explicitly calculated line energies
and oscillator strengths for numerous transitions. For instance, nearly
600 transition lines are included in the oxygen atmosphere model at
$B=10^{12}$ G.

(3) In the case of hydrogen and helium, it is non-trivial to take into
account the effects of a finite nuclear mass \citep{potekhin94,
  bezchastnov98}. However, a simple perturbative approach, introduced
by \citet{pavlov93}, is applicable to mid-Z element atmospheres due to
the larger nuclear mass and binding energies.

(4) All the previous magnetized atmosphere models, except the recent
hydrogen atmosphere models of \citet{potekhin04_2}, adopt
polarization vectors assuming the plasma is fully-ionized. However, such
atmosphere models may not produce correct spectra and absorption
line profiles at low temperatures when the plasma is partially-ionized
\citep{bulik96, potekhin04_2}. In our model, we explicitly compute
polarization vectors by taking into account the effects of bound-bound
and bound-free transitions.

(5) Our radiative transfer code has been extensively tested for
partially-ionized hydrogen atmospheres at various magnetic field
strengths \citep{ho03, potekhin04_2}. The code can be made
applicable to mid-Z element atmospheres in which numerous line and edge
features are present. We solve the coupled radiative transfer equations for
the two photon polarization normal modes, unlike the polarization-averaged
opacities used by \citet{miller92}.

In this paper, we present each step in the construction of mid-Z
element atmosphere models.
We address important physical effects in strongly-magnetized dense plasmas,
following the partially-ionized hydrogen calculations of
\citet{potekhin03}, \citet{ho03} and \citet{potekhin04_2}.
Sections~\ref{sec_atom}, \ref{sec_eos}, and \ref{sec_opacity}
discuss the atomic physics, equation of state, and opacities, respectively,
that we use in the modelling;
more details can be found in \citet{mori02} and \citet{mori06}.
Some aspects of the radiative transfer calculation are given in
Section~\ref{sec_transfer}; details of the radiative transfer scheme are
described in \citet{ho01} and \citet{ho03}.
Section~\ref{sec_profile} shows our atmosphere structure results, and
our model spectra are discussed in Section~\ref{sec_spectra};
we compare our mid-Z element atmosphere models with partially ionized
hydrogen atmosphere models to illustrate the differences
\footnote{Most of the previous NS atmosphere models have been
constructed by assuming a fully-ionized plasma \citep{shibanov92,
zane00, ho01, ozel01, lloyd03}. However, the atomic fraction can be
sizeable for hydrogen atmospheres at the magnetic field strengths and
temperatures typical of the observed INSs \citep{ho03,potekhin04_2}.}.
A discussion on the determination of NS parameters is given in
Section~\ref{sec_ns}.
We compare our model spectra to the observations of INSs
in Section~\ref{sec_obs}.
Section~\ref{sec_summary} summarizes our results.

\section{Atomic physics in strong magnetic field} 
\label{sec_atom} 

In a strong magnetic field, the atomic structure is quite different from
the non-magnetic case. A strong magnetic field deforms the atom into a
cylindrical shape when magnetic field effects are larger than Coulomb
field effects, i.e., at $\beta_Z>1$ (Landau regime), where $\beta_Z=B
/(4.7\times10^{9}Z^2 \mbox{ G})$. A
bound electron in the Landau regime is often denoted by two quantum
numbers $m$ and $\nu$ [hereafter we use $(m\nu)$ to denote a bound
state], where $m$ is the magnetic quantum number and $\nu$ is the longitudinal
quantum number along the field line. $\nu=0$ (tightly-bound) states
have a larger binding energy than $\nu>0$ (loosely-bound) states.
When $\beta_Z\ga1$, the ground state
configuration is composed of bound electrons all in tightly-bound
states, i.e. $(m0)$ states. When the axisymmetry around the magnetic
field is preserved, transitions with $\Delta m = 0$ and $\Delta\nu=odd$
or $\Delta m=\pm 1$ and $\Delta\nu=even$ are allowed by dipole
selection rules. The former transitions occur in the longitudinal
polarization mode ($\alpha=0$, parallel to $B$), while the latter
transitions occur in the circular polarization modes ($\alpha=\pm 1$,
transverse to $B$). Several forbidden transitions (e.g., $\Delta m
= 2$ transition) will have non-zero oscillator strengths when the
motional Stark field breaks the axisymmety around the magnetic field
\citep{potekhin03}. However, we neglect these weak transitions because the
motional Stark effects are significantly smaller in mid-Z element
plasmas.

We developed a fast and accurate atomic code suitable for the Landau
regime \citep{mori02}. The code provides atomic data to better than
1\% and 10\% accuracy for energies and oscillator strengths,
respectively. Most importantly, by applying perturbation methods to
higher Landau levels, we have an extended range of applicable magnetic
fields ($\beta_Z\ga0.3$). The fast algorithm enables us to
compute atomic data for numerous electron configurations within
reasonable CPU time. We also evaluate bound-free cross-sections
following \citet{potekhin97}.

\subsection{Line broadening} 
\label{sec_broad}

Among various line broadening mechanisms in strongly-magnetized mid-Z
element plasmas, motional Stark effects are the most important
\citep{mori06}.  As ions move randomly in a thermal plasma, coupling
of the collective motion and the internal electronic structure induces
a motional Stark field ($\vec{E}_{MS}=\frac{\vec{K}\times\vec{B}}{Mc}$),
where $\vec{K}$ is the pseudomomentum and $M$ is the
mass of the ion.  We evaluated the energy shift caused by the motional
Stark field using the perturbation method
\citep{pavlov93, bezchastnov98} because the atomic mass is much heavier and the
binding energies are significantly larger than in the hydrogen
atom. \citet{pavlov93} introduced a perturbation method for hydrogen
atom while the method of \citet{bezchastnov98} is applicable to
charged ions. In the perturbation method, the second-order energy shift (note that the first-order
term vanishes) is given by $\frac{K^2_\perp}{2M_{\perp,
\kappa}}$, where $K_\perp$ is the pseudomomentum component
perpendicular to the magnetic field and $M_{\perp, \kappa}$ is the
transverse mass ($M_{\perp, \kappa}>M$ ) defined in \citet{pavlov93}
and \citet{bezchastnov98}.  

For a given bound state $\kappa$, one can evaluate $M_{\perp, \kappa}$
from atomic binding energies calculated under the infinite nuclear mass
assumption \citep{pavlov93}.  The increase in the mass due to motional
Stark effects is roughly proportional to $Z^{-2} M^{-1}$ for loosely-bound ($\nu>0$) states.
Therefore, the mass correction factor of the
H-like carbon ion is smaller than that of hydrogen by a factor of
$\sim400$. We computed the transverse mass for bound states with an
electron in a $\nu>0$ state using the atomic data of H-like ions since
bound electrons in a $\nu>0$ state are located at a large distance from
the other bound electrons in tightly-bound states. This procedure saves
a significant amount of computation time, while the results are nearly
identical to those of the more rigorous calculation.  Since the
motional Stark field reduces the binding energy, the line profile has
a wing towards lower energies \citep{pavlov93}. In our opacity calculation,
we computed a line profile for motional Stark broadening following
\citet{pavlov93} and \citet{pavlov95_2}.

Other less important sources of line broadening are thermal Doppler
and pressure broadening.  In a dense plasma, electron bound states are
perturbed by electric fields from electrons (electron collisional
broadening) and ions (quasi-static Stark broadening)
\citep{salzmann98}. We evaluated line widths by the two types of
pressure broadening following \citet{pavlov95_2}.  Pressure broadening
becomes dominant only at high density ($\rho\ga10^2$~g~cm$^{-3}$).

\subsection{Ionization balance} 
\label{sec_ion_balance} 

In local thermodynamic equilibrium (LTE) atmospheres, the ionization fraction
and level population are
determined by the Saha-Boltzmann equilibrium. The validity of the LTE
assumption will be discussed later (see Section~\ref{sec_lte}). We
iteratively solve the
generalized Saha-Boltzmann equations in a magnetic field until
convergence is achieved \citep{khersonskii87, rajagopal97, mori06}.
We take into account pressure ionization by assigning occupation
probabilities for bound states using the electric microfield
distribution of \citet{potekhin02}.
 
\begin{figure}
\resizebox{\hsize}{!}{\includegraphics{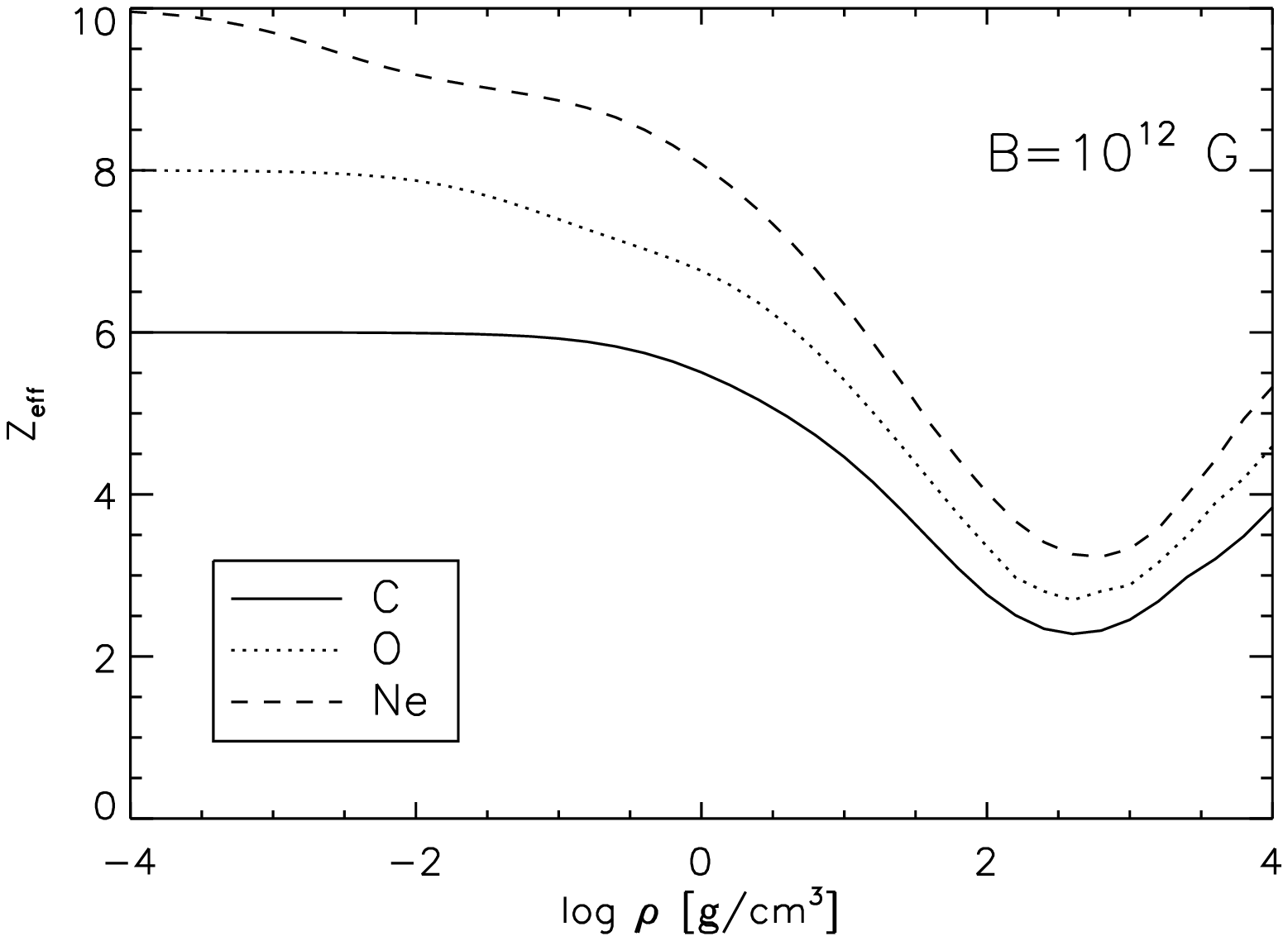}}
\resizebox{\hsize}{!}{\includegraphics{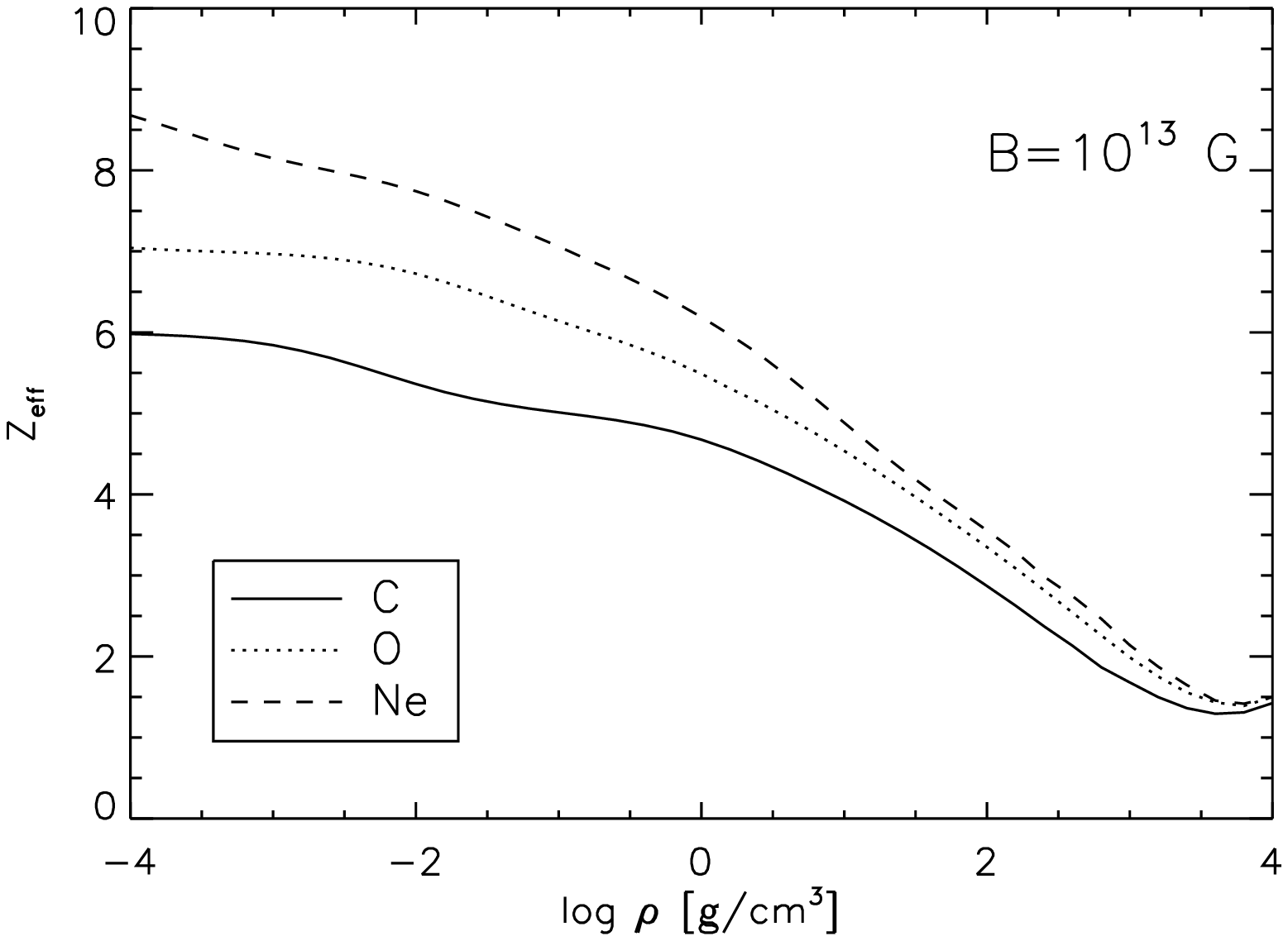}}
\caption{Effective charge $Z_{\rm eff}$ as a function of density 
for carbon, oxygen, and neon plasmas at
$B=10^{12}$~G (top) and $B=10^{13}$~G (bottom) and $T=3\times10^6$~K.
\label{fig_zeff}}
\end{figure}

Figure~\ref{fig_zeff} compares the effective charge ($Z_{\rm eff}$) of
carbon, oxygen, and neon at $10^{12}$~G and $10^{13}$~G. The effective
 charge is defined as $Z_{\rm eff}=\sum_i x_i Z_i$, where $Z_i$ and
 $x_i$ are the charge and the number fraction of ions of the $i$-th
 type ($\sum_i x_i =1$). We adopt
the same temperature ($T=3\times10^6$~K) to illustrate the density
dependence.  In all cases, the degree of ionization decreases with
density until pressure ionization becomes important at high density
(e.g., $\rho\sim10^2$ g~cm$^{-3}$ at $B=10^{12}$~G).
Magnetized hydrogen plasmas exhibit a similar behavior \citep{potekhin99},
though pressure ionization becomes important at
lower densities in hydrogen atmospheres.

The ion fractions have a strong dependence on magnetic field.  The degree
of ionization decreases at higher magnetic fields since the binding energy
increases. Note that an unmagnetized carbon plasma is highly ionized
($Z_{\rm eff}\simeq 5$--6) even at low temperatures
\citep{potekhin05}. At higher magnetic fields, pressure ionization
takes place at higher density because the size of atoms and ions is
smaller.

\begin{figure} 
\resizebox{\hsize}{!}{\includegraphics{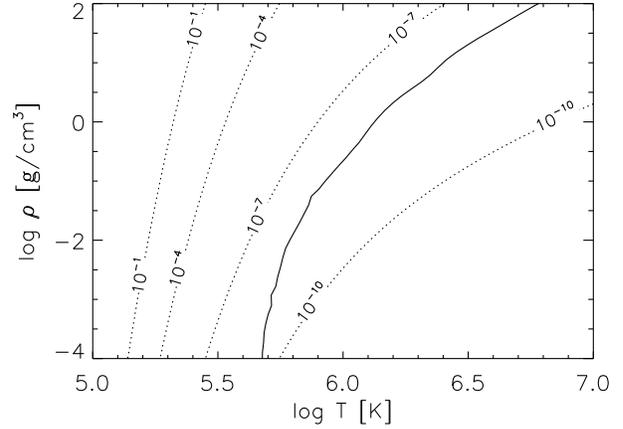}}
\caption{The fractional ratio of the C atom and C$_2$ molecule at
  $B=10^{13}$~G (dotted lines). The solid curve is the carbon
  atmosphere profile with $B=10^{13}$~G and $\Teff=10^6$~K, as discussed in
  Section~\ref{sec_profile}. \label{fig_mol}}
\end{figure}

At high magnetic fields and low temperatures, molecules may be formed in
the atmosphere \citep{lai01}. We find that neutral carbon becomes
abundant at $T\la10^6$~K and $B=10^{13}$~G. Since we do not take into
account any molecular transitions, our models become less accurate in
the regime where molecular contamination is non-negligible.
\citet{medin06_1} calculate binding energies of
non-hydrogenic molecules using density-functional theory. Using their
results, we solve the Saha equation for dissociation equilibrium
between the C atom and C$_2$ molecule \citep{lai97}. Figure~\ref{fig_mol}
shows the fractional ratio of the C atom and C$_2$ molecule at $B=10^{13}$
G. The solid curve is the density-temperature profile of a carbon
atmosphere with $B=10^{13}$~G and $\Teff=10^6$~K
(see Section~\ref{sec_profile}). Despite the fact that the internal
degrees of freedom in molecules (such as ro-vibrational states) can enhance
molecular abundance by up to two orders of magnitude \citep{mori06_3},
we find that molecules can be safely ignored in our study.

\section{Equation of state} 
\label{sec_eos}

Given the results from Section~\ref{sec_ion_balance}, we compute the equation
of state (EOS) of magnetized mid-Z element atmospheres with
\begin{equation} 
P(\rho,
T)=P^e_{id}+P^i_{id}+P^{ee}_{ex}+P^{ei}_{ex}+P^{ii}_{ex}.
\end{equation} 
The first two terms ($P^e_{id}$ and $P^i_{id}$) represent the ideal gas
pressure of electrons and ions. Our calculation of $P^e_{id}$ takes
into account the Landau quantization and electron degeneracy in
a strong magnetic field. In a dense plasma, non-ideal effects are
important, and they can induce negative excess pressure from
electron-electron ($P^{ee}_{ex}$), electron-ion ($P^{ei}_{ex}$) and
ion-ion coupling ($P^{ii}_{ex}$). We compute the excess pressure
following \citet{ichimaru87}, \citet{potekhin99}, and \citet{potekhin03}.

\begin{figure} 
\resizebox{\hsize}{!}{\includegraphics{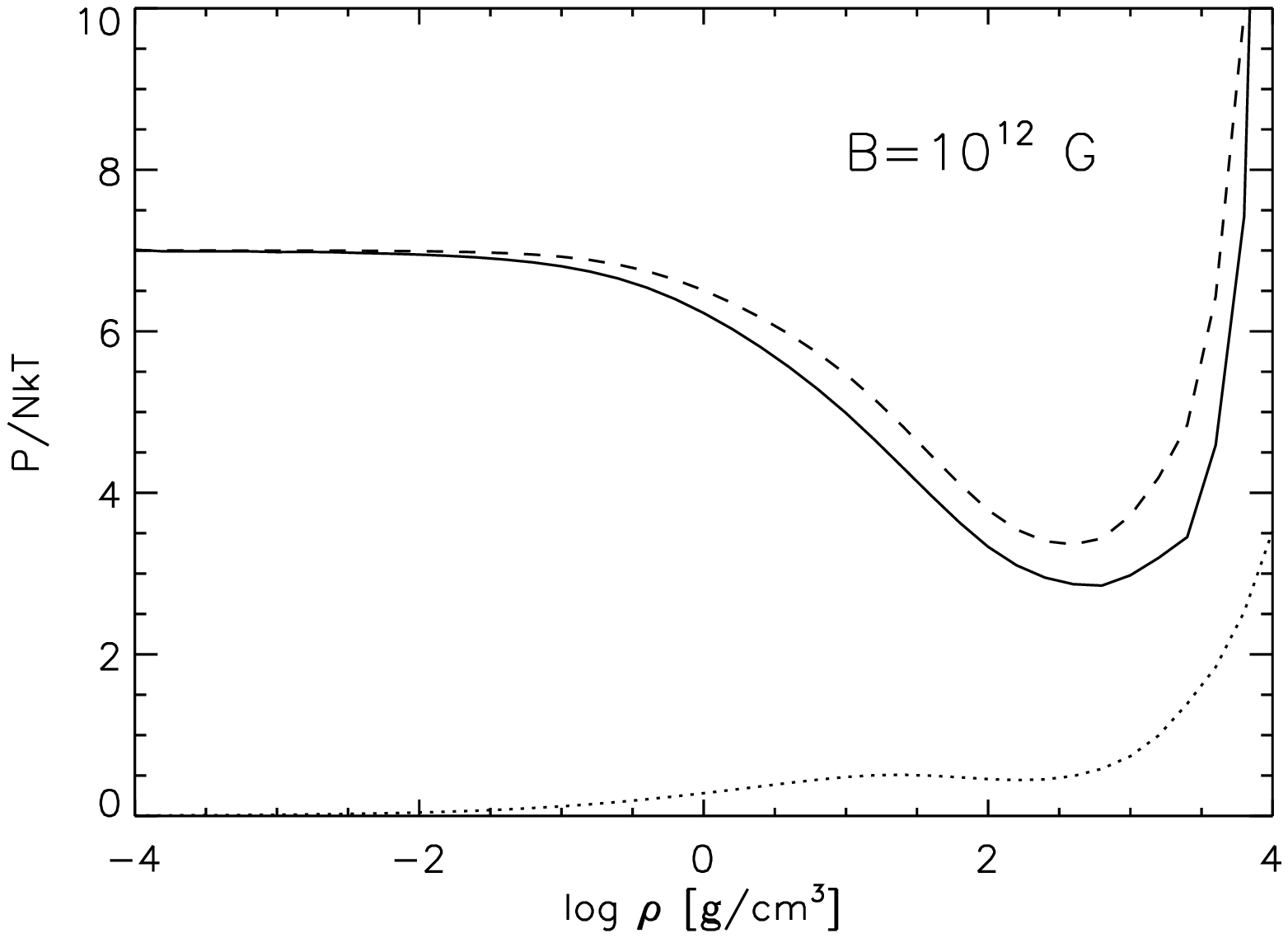}} 
\resizebox{\hsize}{!}{\includegraphics{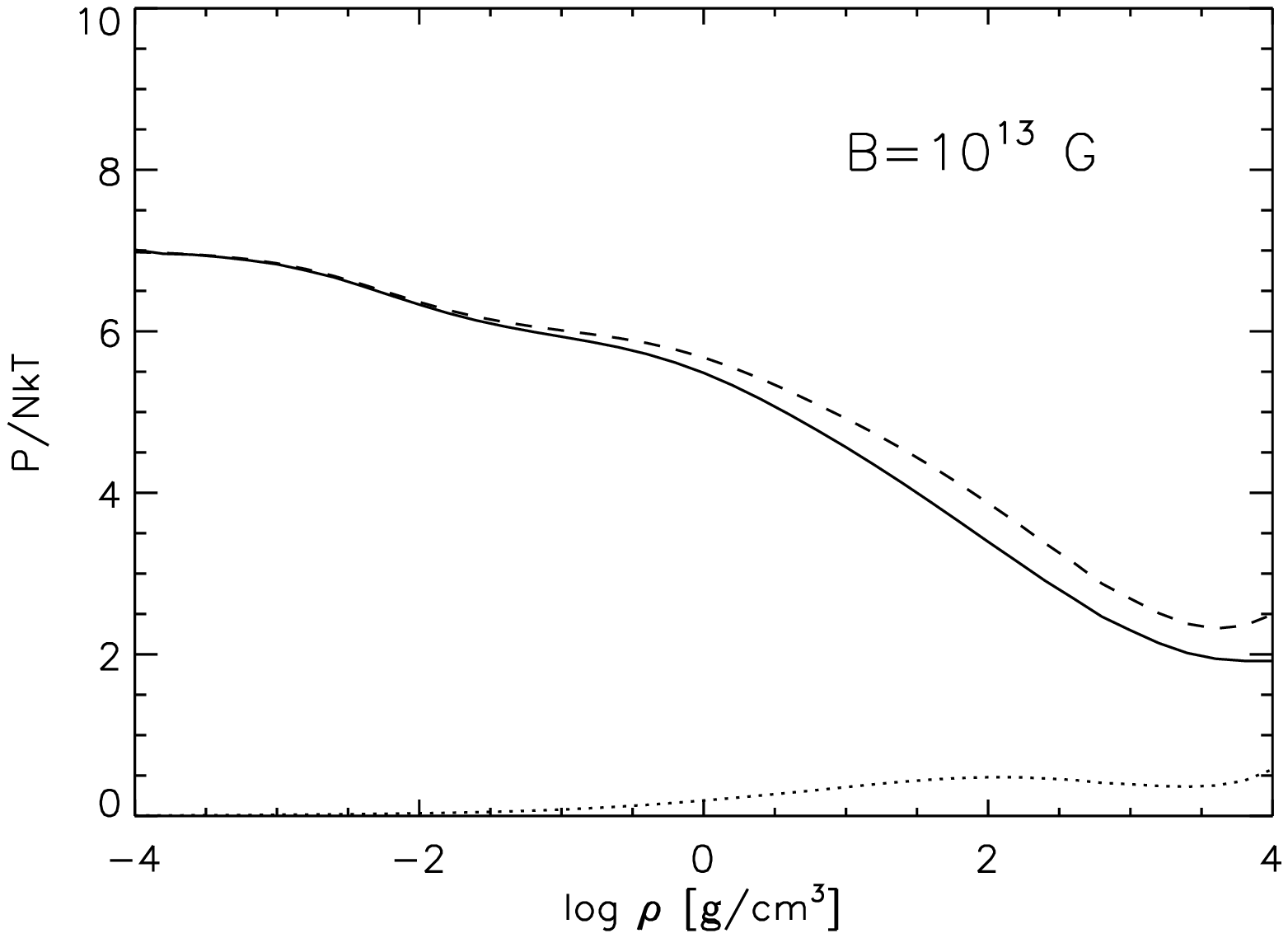}}
\caption{Normalized pressure as a function of density for a carbon plasma
at $B=10^{12}$ G (top)
  and $B=10^{13}$ G (bottom) and $T=3\times10^6$ K. The solid, dashed
  and   dotted are total, ideal (electron+ion) and excess pressure. Note that we plot the magnitude
of the excess pressure since it is negative. \label{fig_pressure}}
\end{figure}

Figure~\ref{fig_pressure} shows the normalized pressure ($P/NkT$, where $N$
is the baryon number density) of a carbon plasma at different magnetic
field strengths and temperatures. We also plot the different
pressure components in Figure~\ref{fig_pressure}.  
The total pressure roughly follows the ion fraction curves in
Figure~\ref{fig_zeff}. The total pressure decreases with density until
the plasma becomes more ionized by pressure ionization. At high
densities (e.g., $\rho\ga10^3$~g~cm$^{-3}$ for $B=10^{12}$~G), electron
degeneracy (dashed line) enhances the pressure sharply although it is
partially compensated by the increasing excess pressure (dotted line).
At higher magnetic fields, such transition points shift to higher
density because the binding energy is increased and the degree of
electron degeneracy is decreased.

For a given magnetic field strength and atmosphere composition, we
produced an EOS table in the temperature range $ 5.4 \le \log_{10}
T \le 7.8$ with a step-size $\Delta \log_{10} T=0.3$ and in the
density range of $-7.0 \le \log_{10}\rho \le 5.0$ with a step-size
$\Delta \log_{10}\rho=1.0$.  The tables extend to low densities in
order to cover strong absorption lines which form in the shallow layers
of the atmosphere.


\section{Opacities} 
\label{sec_opacity}

In this section, we briefly describe our opacity calculation. More
details can be found in \citet{mori06}.  In a strongly-magnetized
plasma, opacity becomes highly anisotropic, and radiation propagates in
two normal modes: the ordinary mode (O-mode), which is mostly polarized
parallel to the magnetic field, and the extraordinary mode (X-mode),
which is mostly polarized perpendicular to the magnetic field
\citep{ginzburg70, meszaros92}. For a given density and temperature,
the opacity (in cm$^{2}$~g$^{-1}$) in the normal mode $j$ is given by
\begin{equation} 
\kappa^j (E, \theta) = \sum_\alpha |\vec{e}_\alpha^j (E, \theta)|^2
\kappa_\alpha(E),
\label{eq_opacity}
\end{equation}
where $|\vec{e}_\alpha^j|$ is the polarization eigenvector;
$|\vec{e}_\alpha^j|$ reflects the dielectric properties of the magnetized
plasma and depends on the angle between the magnetic field and the direction
of photon propagation $\theta$, such that
$\vec{e}_0^j=\vec{e}_z^j$ is the $z$-component (along $B$) of the mode
eigenvector and $\vec{e}_\pm^j=(\vec{e}_x^j\pm i\vec{e}_y^j)/\sqrt{2}$
are the circular components.

\begin{figure} 
 \resizebox{\hsize}{!}{ \includegraphics{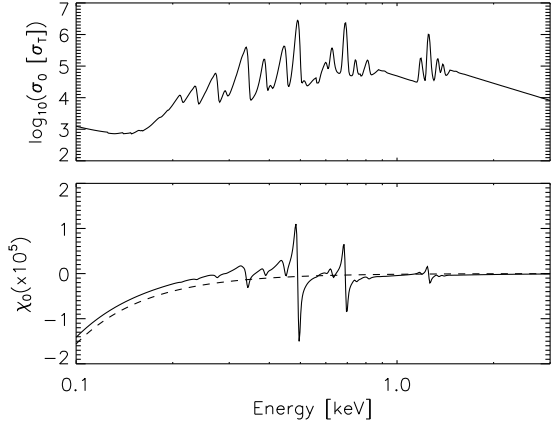}}
\caption{Cross-section (top), in units of Thomson cross-section,
and polarizability tensor (bottom) as a function of energy in the $\alpha=0$
  mode for carbon at $B=10^{12}$~G, $T=5\times10^5$~K and
  $\rho=10^{-2}$~g~cm$^{-3}$. The dashed line shows the polarizability
  tensor for a fully-ionized plasma. The dominant ionization states are
  Li-like and Be-like carbon. \label{fig_chi0}}
\end{figure}

Since mid-Z element atmospheres are usually not fully-ionized, we
include the effects of bound species by using of the Kramers-Kronig
relation \citep{bulik96, potekhin04_2}. Figure~\ref{fig_chi0} shows
the cross-section and polarizability tensor in the $\alpha=0$ mode
when carbon is partially-ionized ($Z_{\rm eff}=2.9$). The cross-section
($\sigma_0$) is dominated by bound-bound and bound-free transitions in
the X-ray band. The polarizability tensor ($\chi_0$) deviates
significantly from that in a fully-ionized plasma, particularly near the
strong lines.

\begin{figure} 
\resizebox{\hsize}{!}{  \includegraphics{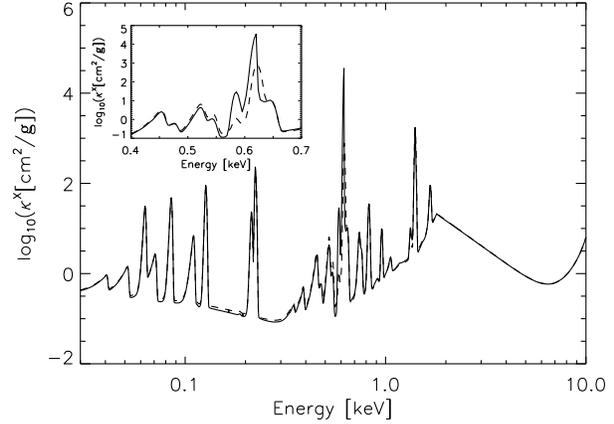}}
\caption{X-mode opacity of carbon at $B=10^{12}$~G,
 $T=2\times10^6$~K, $\rho=10^{-2}$~g~cm$^{-3}$, and $\theta=10^\circ$.
  The solid and
  dashed lines show the X-mode opacity for the complete and hybrid models,
  respectively. The fraction of bare and H-like carbon is 81 and 19\%,
  respectively, while other ionization species have fractions less than
  0.01\%. The inset is a magnified view of the
  $E=0.4$--0.7~keV region to illustrate the difference between the
  complete and hybrid models in the vicinity of the strongest
  bound-bound transition line at $\sim 0.6$~keV.  \label{fig_opacity}}
\end{figure}

When the degree of ionization increases, the difference between
using the polarizability tensors for partial ionization (complete
model) and full ionization (hybrid model) becomes smaller.
Figure~\ref{fig_opacity} shows the X-mode opacity of carbon at
$B=10^{12}$~G in the complete and hybrid models for a nearly fully-ionized
case ($Z_{\rm eff}=5.8$). Although the overall difference is tiny, there
are noticeable deviations in the vicinity of the strong bound-bound
transition lines at $E\sim0.6$~keV (see inset).
Our results indicate that the strength and shape of strong
absorption lines can be strongly affected even when the plasma is nearly
fully-ionized.

We produced opacity tables for the same temperature and density grids
as the EOS tables. For a given set of temperature and density, our
opacity tables consist of cross-sections $\sigma_\alpha(E)$ and
polarizability tensors $\chi_\alpha(E)$ in the three cyclic
polarization modes ($\alpha=0, \pm1$).  We adopted $\sim2500$
equal logarithmically-spaced energy grid points; this allows us to resolve
narrow lines.


\section{Radiative transfer} 
\label{sec_transfer}

Thermal radiation from a NS is mediated by an atmosphere (with scale
height $\sim 1$~cm) that covers the stellar surface.
To determine the emission properties of a magnetized
atmosphere, the radiative transfer equations for the two coupled
photon polarization modes are solved \cite[see][for details on the
construction of the atmosphere models]{ho01, ho03}.
We note here a few particulars of our models.

The EOS tables for a given magnetic field are arranged by temperature
and density.  The pressure at Thomson depth $\tau$
[$=-\keso\int\rho\,dz$, where $\keso=(Z/A)\sigth/\mpr$, the atomic mass
number is $A$, and the Thomson cross-section is $\sigth$] is calculated from
hydrostatic equilibrium.  With a given temperature profile, we search
the EOS tables for the nearest temperatures and gas pressures and do a
weighted average to obtain the density profile.  To include radiation
pressure $P_{rad}$, we calculate $P_{rad}$ from its contribution
to hydrostatic equilibrium \citep{mihalas78}
\begin{equation} 
\frac{dP_{rad}}{d\tau} = \frac{\sigsb\Teff^4}{c}\frac{\kfmean}{\keso},
\end{equation}
where
\begin{equation}
\kfmean = \frac{1}{F}\sum_j\int_0^\infty d\nu\int d\Omega\,
 \kappa^{\rm tot}_j F_j,
\end{equation}
$F_j$ is the specific flux for mode $j$, and $F$ is total flux;
this contribution is usually small (see Section~\ref{sec_radpres}).
The radiation pressure is then subtracted from the total pressure to
obtain the gas pressure.

The self-consistency of the atmosphere model is determined by
requiring that the fractional temperature corrections $\Delta
T(\tau)/T(\tau)\ll 1\%$ at each Thomson depth, deviations from
radiative equilibrium $\ll 1\%$, and deviations from constant total flux
$< 1\%$.  Note that the atmosphere models formally have a dependence,
through hydrostatic balance, on the surface gravity
$g$~$[=(1+\zg)GM/R^2]$ and thus the NS mass $M$ and radius $R$; however,
the emergent
spectra do not vary much using different values of $g$ around $2\times
10^{14}$~cm~s$^{-2}$
\citep[see Section~\ref{sec_mag_grav}; see also][]{pavlov95_1}.
Most of our models are constructed using a
surface gravity $g=2.4\times 10^{14}$~cm~s$^{-2}$ with $M=1.4M_\odot$,
$R=10$~km, and $\zg=0.3$ (see Section~\ref{sec_mag_grav} for other cases).
Also, though our atmosphere models can have a magnetic field at an
arbitrary angle $\Theta_B$ relative to the surface normal, most of the
models considered in Section~\ref{sec_spectra} have the magnetic field
aligned perpendicular to the stellar surface ($\Theta_B=0^\circ$;
see Sections~\ref{sec_mag_grav} and \ref{sec_bfield} for other cases).
The spectra represent emission from a local patch of the NS surface.

\section{Atmosphere profile} 
\label{sec_profile} 

\begin{figure} 
\resizebox{\hsize}{!}{  \includegraphics{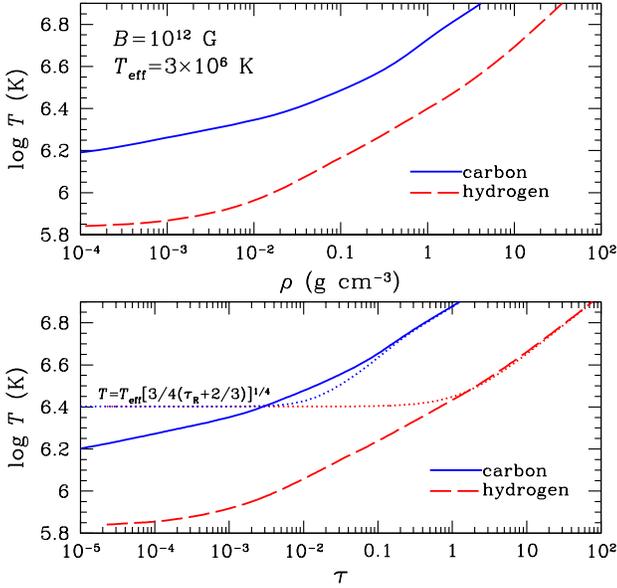}}
\caption{Temperature profiles of carbon (solid lines) and hydrogen
atmospheres (dashed lines) with $B=10^{12}$~G and $\Teff=3\times10^6$~K.
The dotted lines are grey profiles. $\tau$ is the Thomson depth
and $\tau_{\rm R}$ is the Rosseland depth.  At large $\tau$, temperature
profiles merge with grey profiles. \label{fig_t_profile12}}
\end{figure}

\begin{figure} 
 \resizebox{\hsize}{!}{ \includegraphics{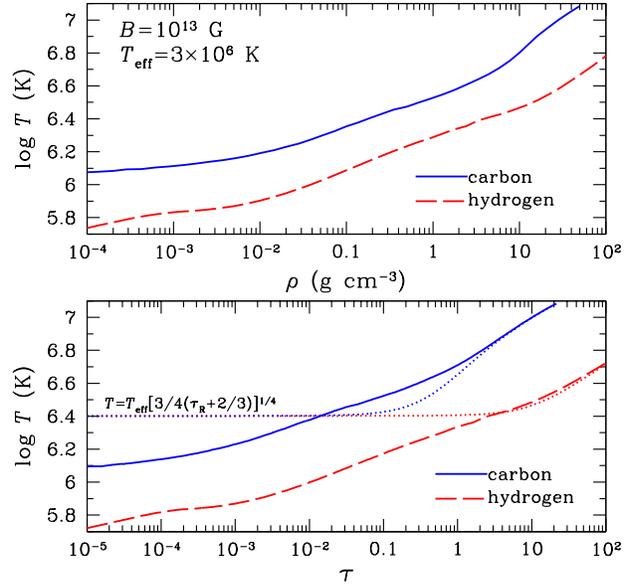} }
\caption{Temperature profiles of carbon and hydrogen with $B=10^{13}$~G and
$\Teff=3\times10^6$~K. Notation is the same as in
Figure~\ref{fig_t_profile12}.  \label{fig_t_profile13}}
\end{figure}

Figures~\ref{fig_t_profile12} and \ref{fig_t_profile13} show the
temperature profiles of carbon atmospheres in comparison with those of
partially ionized hydrogen atmospheres at $B=10^{12}$~G and $B=10^{13}$~G.
There are two distinct differences between hydrogen and mid-Z element
atmosphere profiles; these differences are due to the fact that
bound-bound and bound-free opacities are more important than free-free
opacities. Temperature profiles in the mid-Z element atmospheres are
closer to grey than in hydrogen atmospheres since the opacities are
less energy-dependent, while the free-free opacity (which is dominant in
hydrogen atmospheres) has a steep energy dependence.  Photosphere densities
are also significantly smaller since opacities are higher than in hydrogen
atmospheres, and photosphere densities increase with magnetic field since
opacities are reduced.

\begin{figure} 
 \resizebox{\hsize}{!}{ \includegraphics{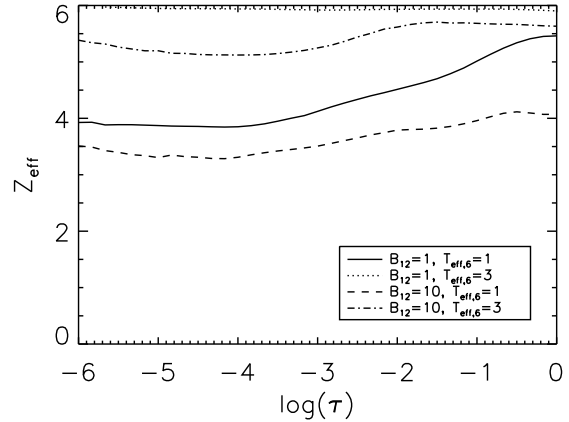}}
\caption{Effective charge of carbon as a function of Thomson depth $\tau$
at $B=10^{12}$~G and $B=10^{13}$~G and $\Teff=10^6$ and $3\times10^6$~K.
Carbon is nearly fully-ionized at $\Teff=5\times10^6$~K. \label{fig_c_frac}}
\end{figure}

\begin{figure} 
 \resizebox{\hsize}{!}{ \includegraphics{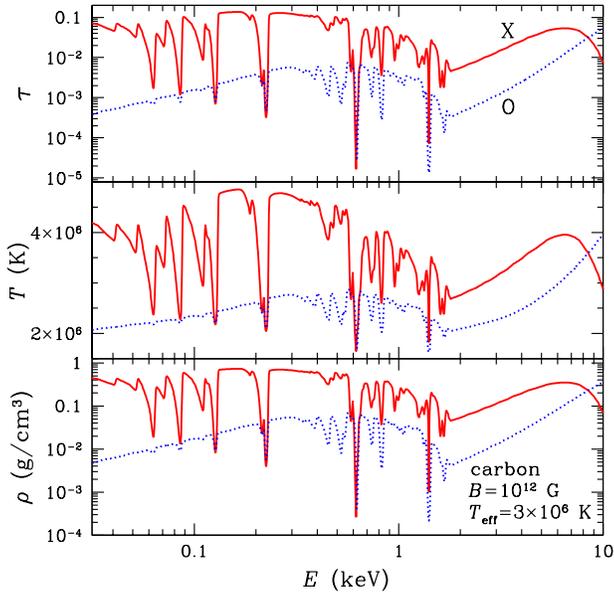}}
\caption{Thomson depth (where photons decouple from matter) as a
function of energy for a carbon atmosphere with $B=10^{12}$~G and
$\Teff=3\times10^6$~K. The bottom two panels plot the local
temperature and density at the corresponding Thomson depth.  The
solid and dotted lines are for the X-mode and O-mode, respectively.
For the two polarization modes, the photon-decoupling layer is defined
such that the effective (angle-averaged) optical depth of each mode is unity.
See Section~4.2 in \citet{ho01} for more details on how we
calculate effective optical depths. \label{fig_opdepth}}
\end{figure}

Figure~\ref{fig_c_frac} shows the effective charge of carbon atmospheres at
different optical depths. At $B=10^{12}$~G and $\Teff\ga3\times10^6$~K,
the H-like carbon ion is largely populated. At higher $B$ and
lower $\Teff$, various ionization states exist in the atmosphere, and they
are likely to produce numerous spectral features. Figure~\ref{fig_opdepth}
illustrates where X-mode and O-mode photons decouple
from the matter as a function of energy. It is clear that X-mode
photons come from deeper, hotter layers in the atmosphere than
O-mode photons; therefore spectra are strongly polarized and dominated
by the X-mode photons.

\subsection{Radiative levitation}
\label{sec_radpres}

At high temperature, mid-Z element ions may be levitated by radiation
pressure through strong absorption lines. Our scheme described in Section
\ref{sec_transfer} does not perform radiative transfer of individual
lines since we compute the gradient
of radiation pressure using opacities that are integrated over
frequency. Although detailed line-by-line radiative transfer
modeling is beyond the scope of this paper, we estimate here the effects
of radiative levitation in mid-Z element
atmospheres.   
Gravitational force on a carbon
ion is $4.6\times10^{-9}$~dyne when $g=2.4\times10^{14}$~cm~s$^{-2}$,
while radiation force through an
absorption line is $1.6\times10^{-11} \sigma_{l}(T_{eff}/10^7\mbox{ K})^4$~dyne, where
$\sigma_{l}$ is the line cross-section in units of Thomson cross-section
($\sigma_{T}$). In the carbon atmosphere at 
$B=10^{12}$~G and $\Teff=5\times10^6$~K, several strong
lines have cross-sections as large as $\sim10^4\sigma_{T}$. Therefore,
radiative levitation can be important through strong absorption lines
at $\Teff\sim5\times10^{6}$~K, while it will be negligible at lower
temperatures. 

\subsection{Validity of LTE assumption} 
\label{sec_lte}

The LTE assumption may become invalid at low density and high
temperature. To our knowledge, the validity of LTE assumption has not
been investigated in previous atmosphere models for cooling
NSs. Three conditions need to be fulfilled for a LTE plasma
\citep{salzmann98}: (1) Maxwell-Boltzmann distribution of the electron
and ion velocities, (2) Saha equilibrium for the charge state
distribution, and (3) Boltzmann distribution of the excited states. The
first condition is well-satisfied in our case since the electron
self-collision time is very short. However, the latter two conditions
are usually more restrictive. We examine the second condition
following \citet{salzmann98}. For the third condition, we estimate
radiative and collisional transition rates between the ground and
excited states and apply the criteria from \citet{griem64}.

\begin{figure} 
\resizebox{\hsize}{!}{\includegraphics{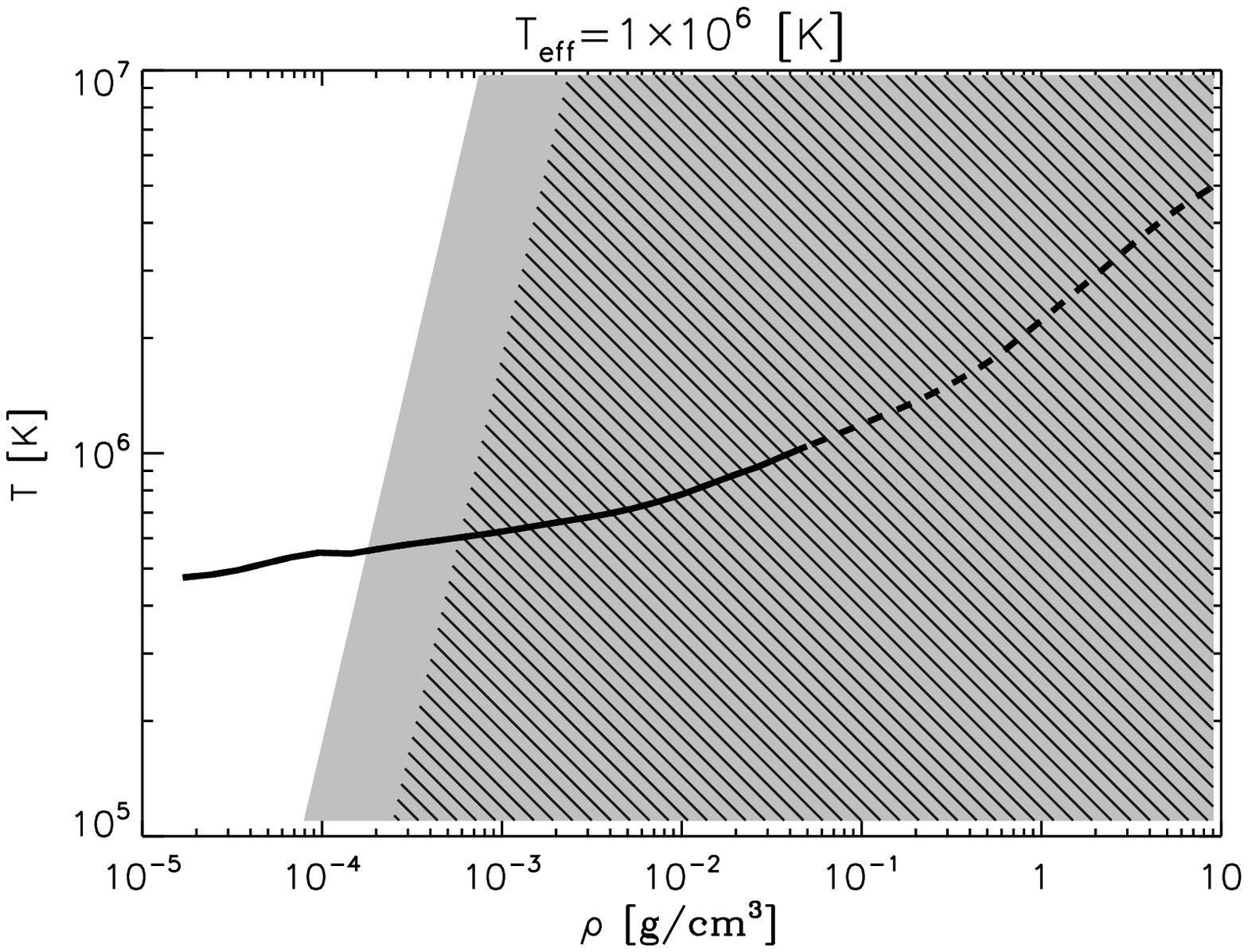}} 
\resizebox{\hsize}{!}{\includegraphics{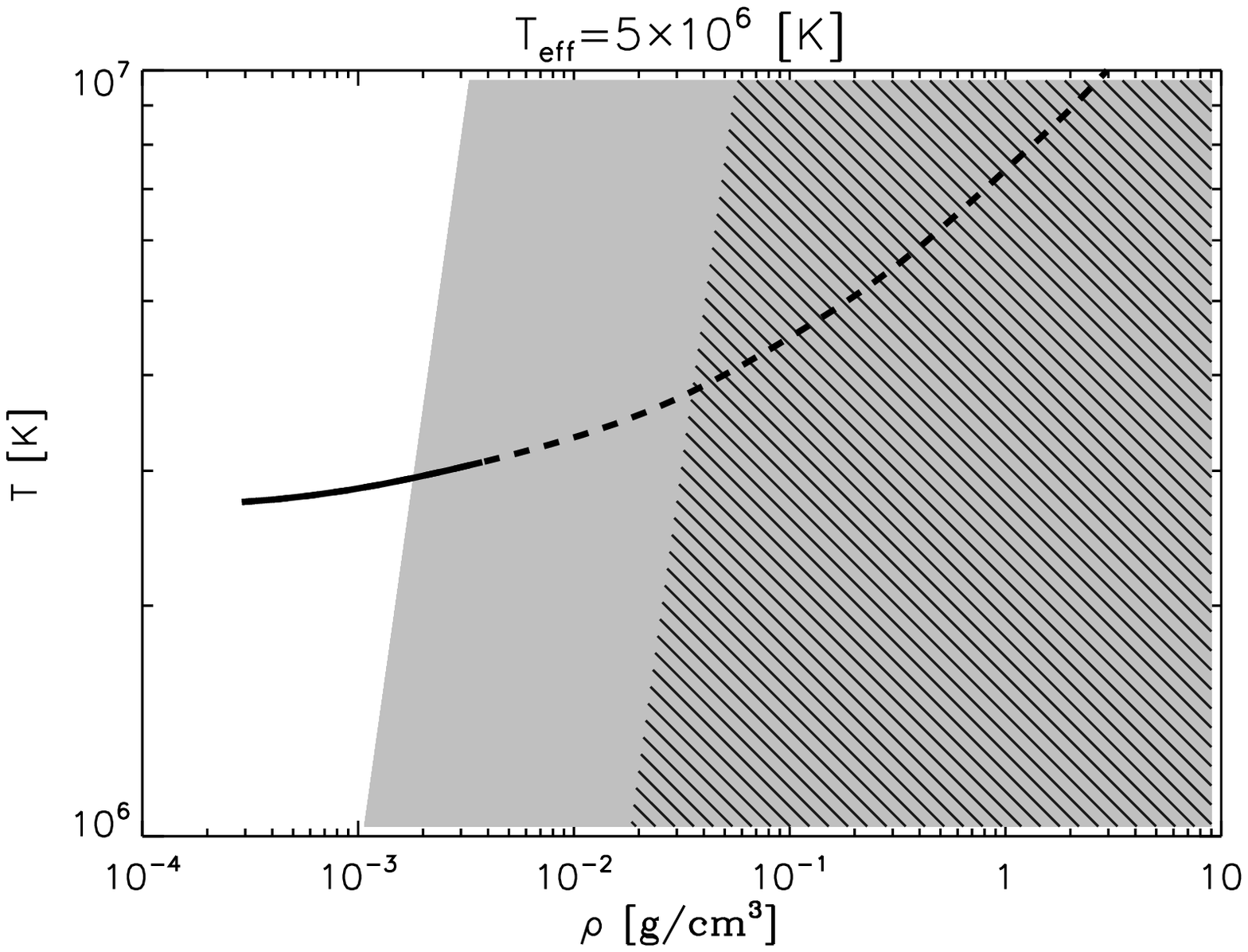}}
\caption{ LTE conditions for carbon atmospheres at $B=10^{12}$~G and
$\Teff=10^{6}$~K (top) and
$\Teff=5\times10^6$~K (bottom). The Saha equation is valid in the
grey area, while the Boltzmann distribution for the excited states is
valid in the hatched area. The solid + dashed curves show the temperature
profile of the atmospheres. The fraction of excited states exceeds 10\% on
the dashed
lines, while the occupation probability of the ground state is larger
than 90\% on the solid lines.
\label{fig_lte}} 
\end{figure}

In Figure~\ref{fig_lte}, we plot the $(\rho, T)$ regime where the LTE
assumption is valid, along with carbon atmosphere profiles at
$B=10^{12}$~G and two effective temperatures. The condition
for a Boltzmann distribution (hatched area) is more restrictive than
the one for Saha equilibrium (grey area). We find that both
conditions break down in the upper atmosphere layers with
$\rho\la10^{-4}$~g~cm$^{-3}$ and $10^{-3}$~g~cm$^{-3}$ at $\Teff=10^6$~K
and $5\times10^6$~K, respectively. Therefore, only a few of the strongest
lines will be affected by non-LTE effects. At higher densities, Saha
equilibrium is recovered. The condition for the Boltzmann distribution
(hatched area) becomes irrelevant in the ($\rho, T$) regime where a
fraction of excited states is small (solid lines). Therefore, the LTE
assumption is completely valid above $\rho\sim10^{-4}$~g~cm$^{-3}$ at
$\Teff=10^6$~K. On the other hand, the occupation probabilities of
excited states 
computed by using the Boltzmann distribution may not be correct in the
region between $\rho\sim4\times10^{-3}$ and $3\times10^{-2}$~g~cm$^{-3}$
at $\Teff=5\times10^6$~K. We find that such a region does
not exist at lower effective temperatures.

We thus expect that only a few strong absorption lines will be
altered by non-LTE effects, while the bulk of the mid-Z element
atmosphere structure and spectra are unaffected in our models.
To address the non-LTE effects, one must find
collisional-radiative steady states by solving rate equations with
both radiative and collisional processes included \citep{salzmann98}. Such
an analysis is beyond the scope of this paper.


\section{Model atmosphere spectra} 
\label{sec_spectra}

\begin{figure} 
 \resizebox{\hsize}{!}{\includegraphics{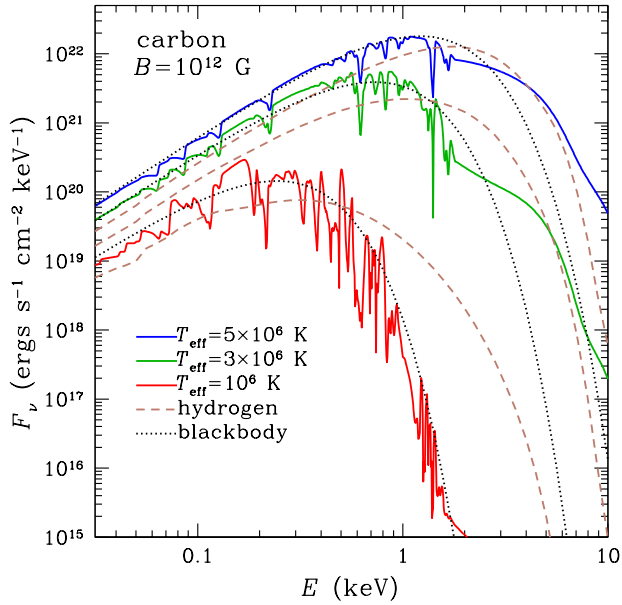}}
\caption{Spectra of carbon atmospheres with $B=10^{12}$~G and three
effective temperatures ($\Teff=1, 3$ and $5\times10^6$~K).
The dashed lines show partially ionized hydrogen atmosphere spectra
with the same $B$ and
$\Teff$, and the dotted lines show blackbody spectra with $T=\Teff$.
\label{fig_spec_c12}} 
\end{figure}

\begin{figure} 
 \resizebox{\hsize}{!}{\includegraphics{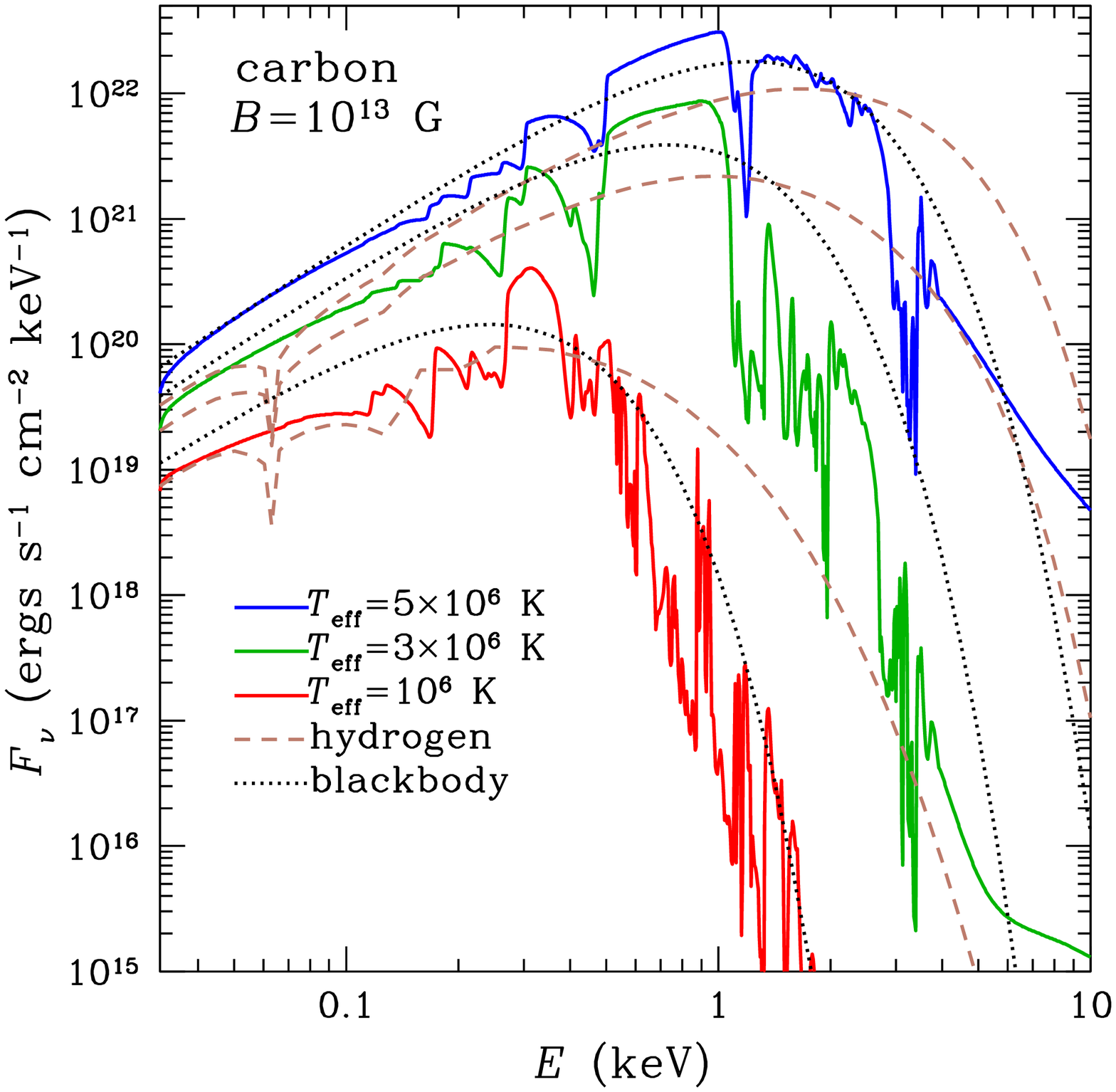}}
\caption{Spectra of carbon atmospheres with $B=10^{13}$~G.
Notation is the same as in Fig.~\ref{fig_spec_c12}.
  \label{fig_spec_c13}} 
\end{figure}

\begin{figure} 
 \resizebox{\hsize}{!}{\includegraphics{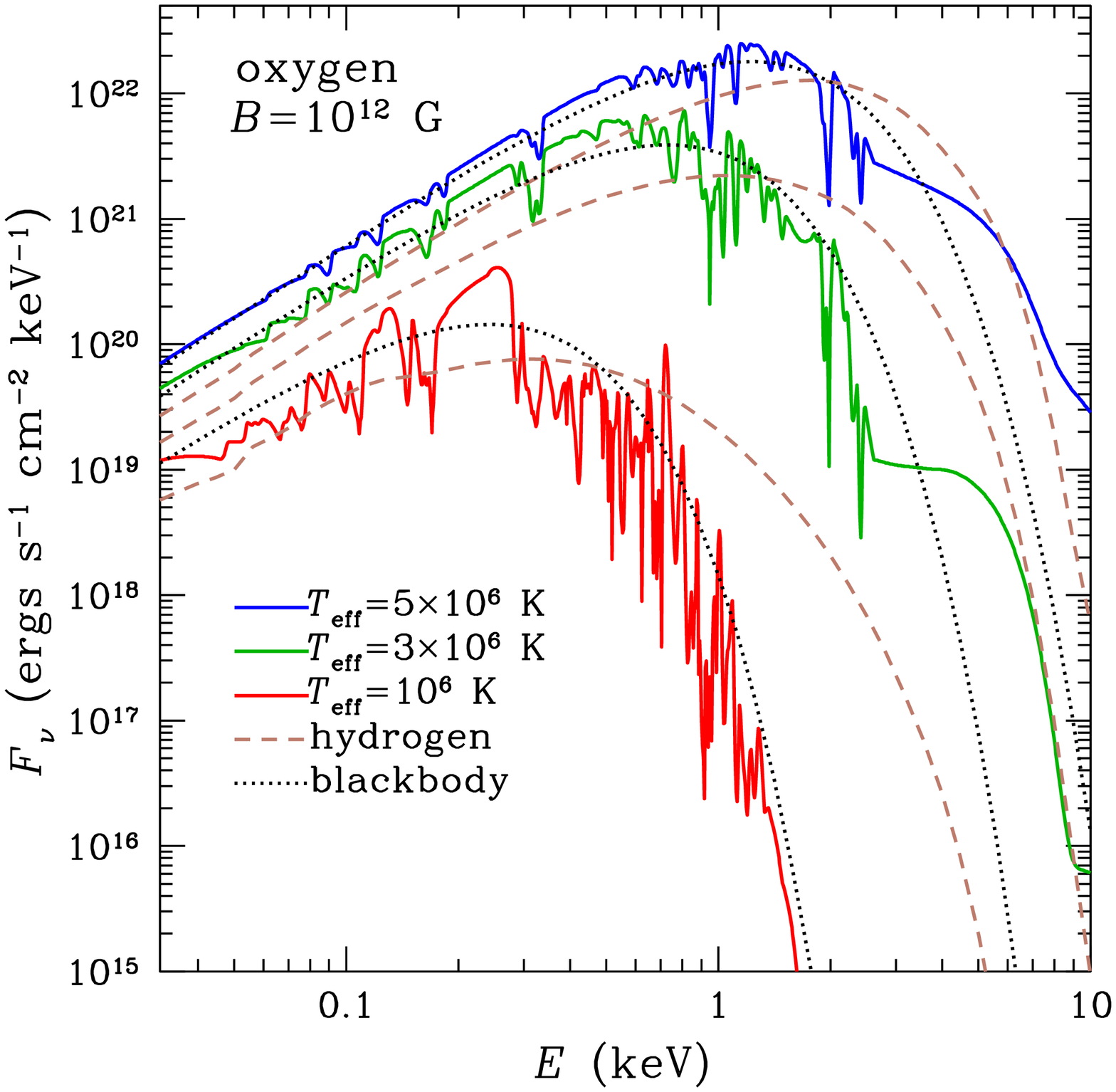}}
\caption{Spectra of oxygen atmospheres with $B=10^{12}$~G.
Notation is the same as in Fig.~\ref{fig_spec_c12}.
 \label{fig_spec_o12}}
\end{figure}

\begin{figure} 
 \resizebox{\hsize}{!}{\includegraphics{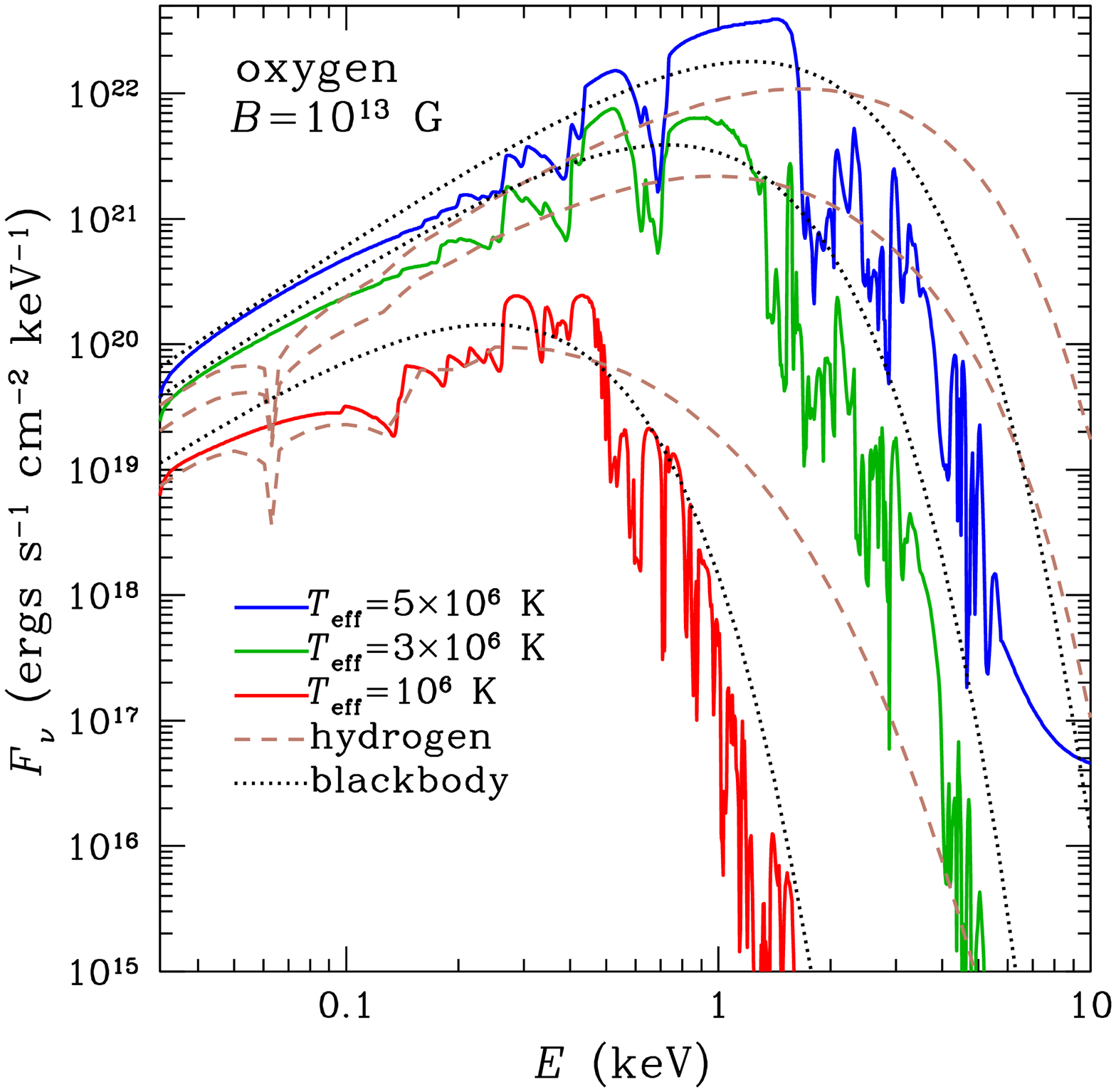}}
\caption{Spectra of oxygen atmospheres with $B=10^{13}$~G.
Notation is the same as in Fig.~\ref{fig_spec_c12}.
 \label{fig_spec_o13}}
\end{figure}

\begin{figure} 
 \resizebox{\hsize}{!}{\includegraphics{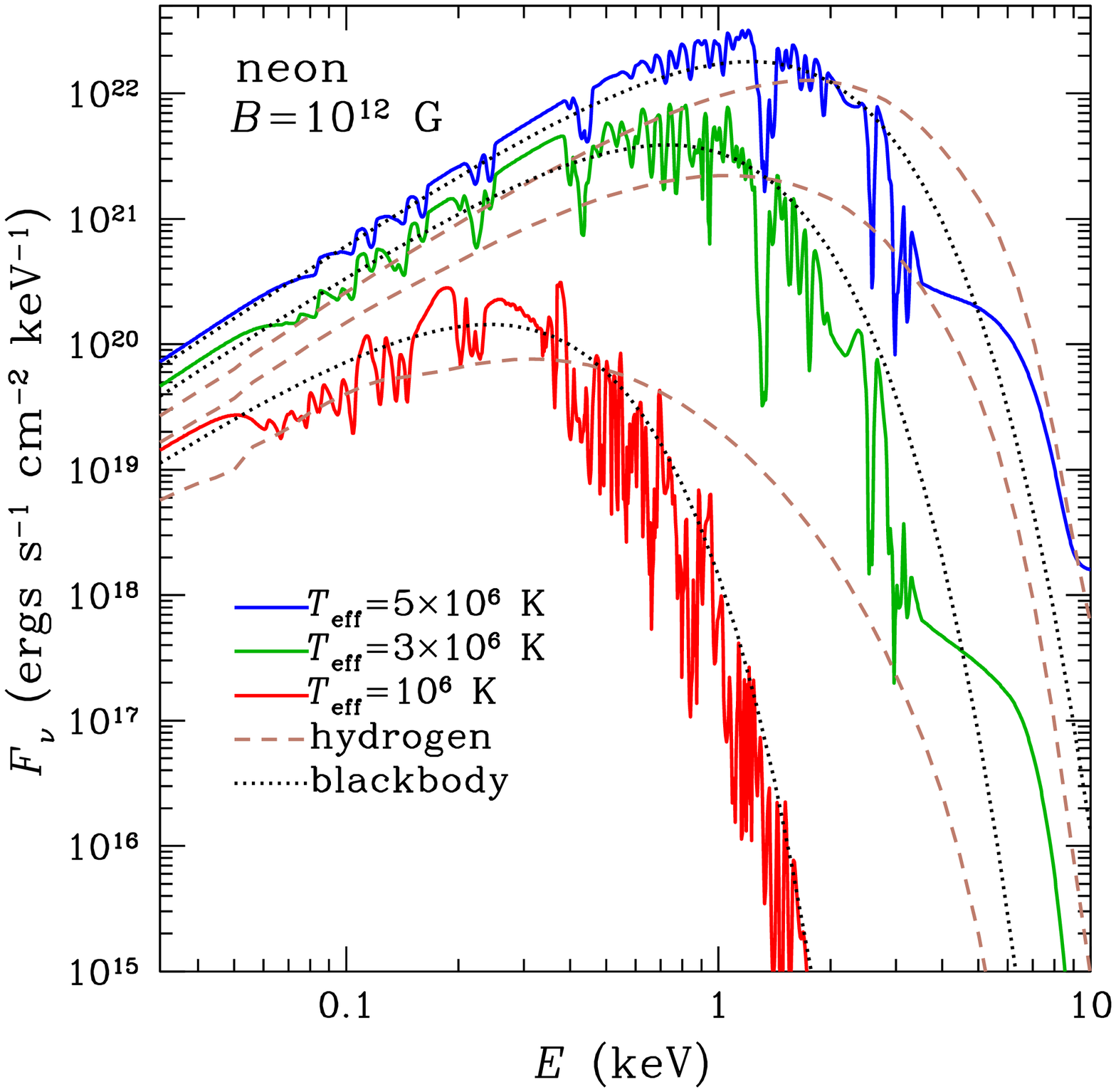}}
\caption{Spectra of neon atmospheres with $B=10^{12}$~G.
Notation is the same as in Fig.~\ref{fig_spec_c12}.
\label{fig_spec_ne12}}
\end{figure}

\begin{figure} 
 \resizebox{\hsize}{!}{\includegraphics{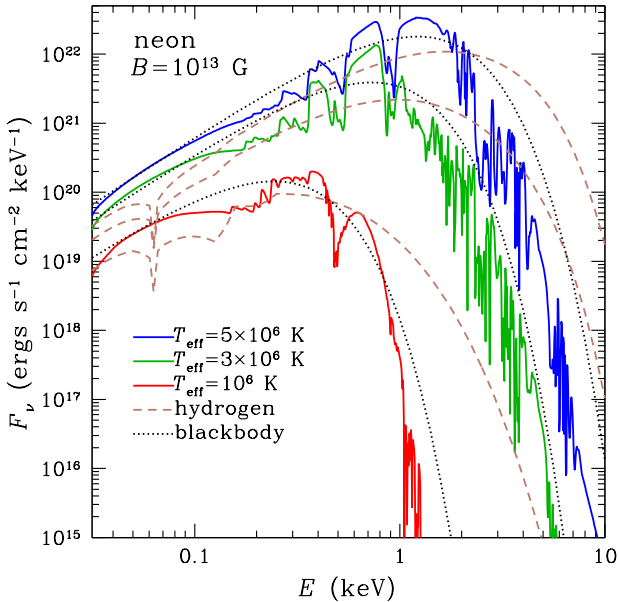}}
\caption{Spectra of neon atmospheres with $B=10^{13}$ G.
Notation is the same as in Fig.~\ref{fig_spec_c12}.
\label{fig_spec_ne13}}
\end{figure}

We construct spectra of mid-Z element atmospheres with
$B=10^{12}$~G and $10^{13}$~G and three different effective temperatures
($\Teff=10^6, 3\times10^6$, and $5\times10^6$~K). The results are
shown for carbon (Figures~\ref{fig_spec_c12} and \ref{fig_spec_c13}),
oxygen (Figures~\ref{fig_spec_o12} and \ref{fig_spec_o13}) and neon
(Figure~\ref{fig_spec_ne12} and \ref{fig_spec_ne13}).
For comparison, we plot partially ionized hydrogen atmosphere spectra
(dashed lines) and blackbody spectra
(dotted lines) with the same magnetic field strengths and temperatures.

\subsection{Spectral energy distribution}

Overall, mid-Z element atmosphere spectra are significantly softer than
hydrogen atmosphere spectra (dashed lines) and close to blackbody spectra
(dotted lines) because temperature profiles are closer to
grey. The mid-Z element atmosphere spectra are even softer than
blackbody in the Wien tail due to the photo-absorption edges (at $\sim
1.5-2$~keV for $B=10^{12}$~G models). The absorption edges are due to
bound-free transitions from the innermost electron [$(m\nu)=(00)$
state] in different ionization species. At higher energy, the spectra
become harder than blackbody and hydrogen atmospheres in order to maintain radiative
equilibrium. A similar feature is seen in the magnetized iron
atmosphere models of \citet{rajagopal97}. This appears to be a common
signature of heavy element atmospheres over a large range of $B$ and
$\Teff$.
 
\subsection{Absorption features}

Numerous absorption lines are present, especially in low temperature
models, and as expected, heavier element atmospheres show more
absorption features.  Magnetic field
dependence is also remarkable. There are more discrete absorption lines at
$B=10^{12}$~G than at $B=10^{13}$~G. Most of the absorption lines at
$B=10^{12}$~G are longitudinal transitions ($\Delta m=0, \Delta \nu>0$ in the
$\alpha=0$ mode). On the other hand, the londitudinal transitions become significantly
weaker at higher $B$ because the $\alpha=0$ component of the polarization
vector components in the X-mode becomes smaller with increasing
$B$. This effect is most noticeable in the low temperature models (e.g., see
the neon models at $\Teff=10^6$~K). The suppression of
longitudinal transitions is also seen in magnetized hydrogen
atmospheres \citep{ho03}. Note that longitudinal transitions appear
strong in the mid-Z element atmospheres at $B=10^{12}$ G since
the magnetic field is not as dominant over the nuclear Coulomb
field at $\beta_Z\ga1$. 

In the $B=10^{13}$~G model spectra, it is mainly the transverse transition
lines that are visible.  At $B=10^{13}$~G, $\Delta m=+1, \Delta
\nu=0$ transitions from bound electrons in outer shells are 
strong (e.g., lines at $\sim0.26, 0.39, 0.69$ and 1.8~keV in oxygen atmosphere models at
$B=10^{13}$ G). The lines appear at lower energies when the temperature is lower and
oxygen is less ionized (compare oxygen spectra at $\Teff=10^6$~K and
$3\times10^6$~K for $B=10^{13}$~G in Figure~\ref{fig_spec_o13}).
Other weaker and broader transverse transition
lines ($\Delta m=\pm1$, $\Delta \nu>0$) are also present in the
$B=10^{13}$~G model spectra.

Most absorption lines have asymmetric profiles with low energy tails
due to motional Stark effects. Some of the strong lines show more
complex profiles due to the enhanced polarizability near the lines. In
general, line widths become larger at higher magnetic fields since motional
Stark effects are more important. Only photo-absorption edges from the 
innermost electrons are clearly visible at high energy where there are no
bound-bound transition lines (e.g., $E\ga1.7$ keV for carbon
atmosphere spectra at $B=10^{12}$ G), although bound-free transitions
are important throughout the X-ray band.

\subsection{Dependence on magnetic field geometry and surface gravity}
 \label{sec_mag_grav}

\begin{figure} 
\resizebox{\hsize}{!}{  \includegraphics{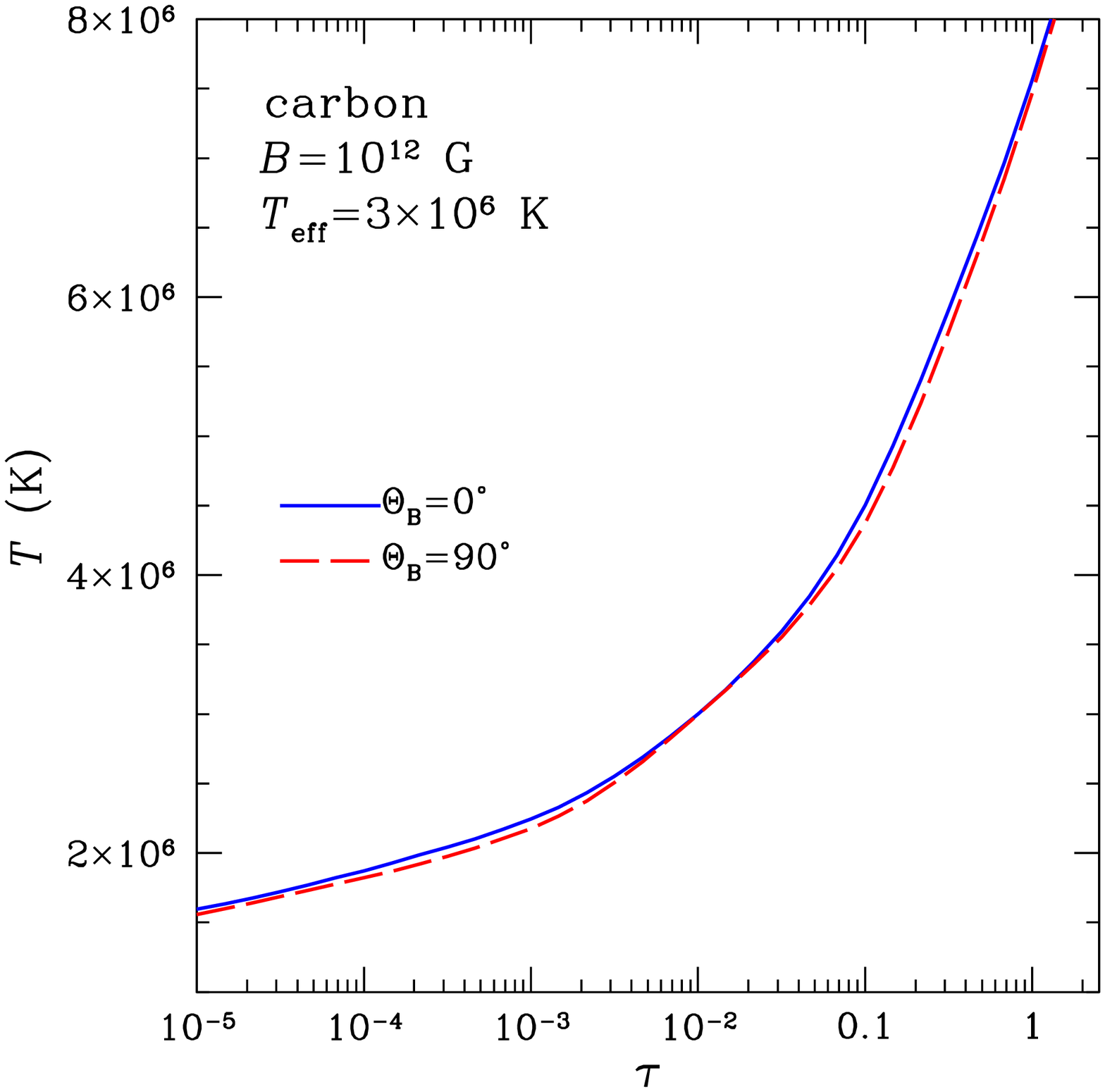}}  
\resizebox{\hsize}{!}{\includegraphics{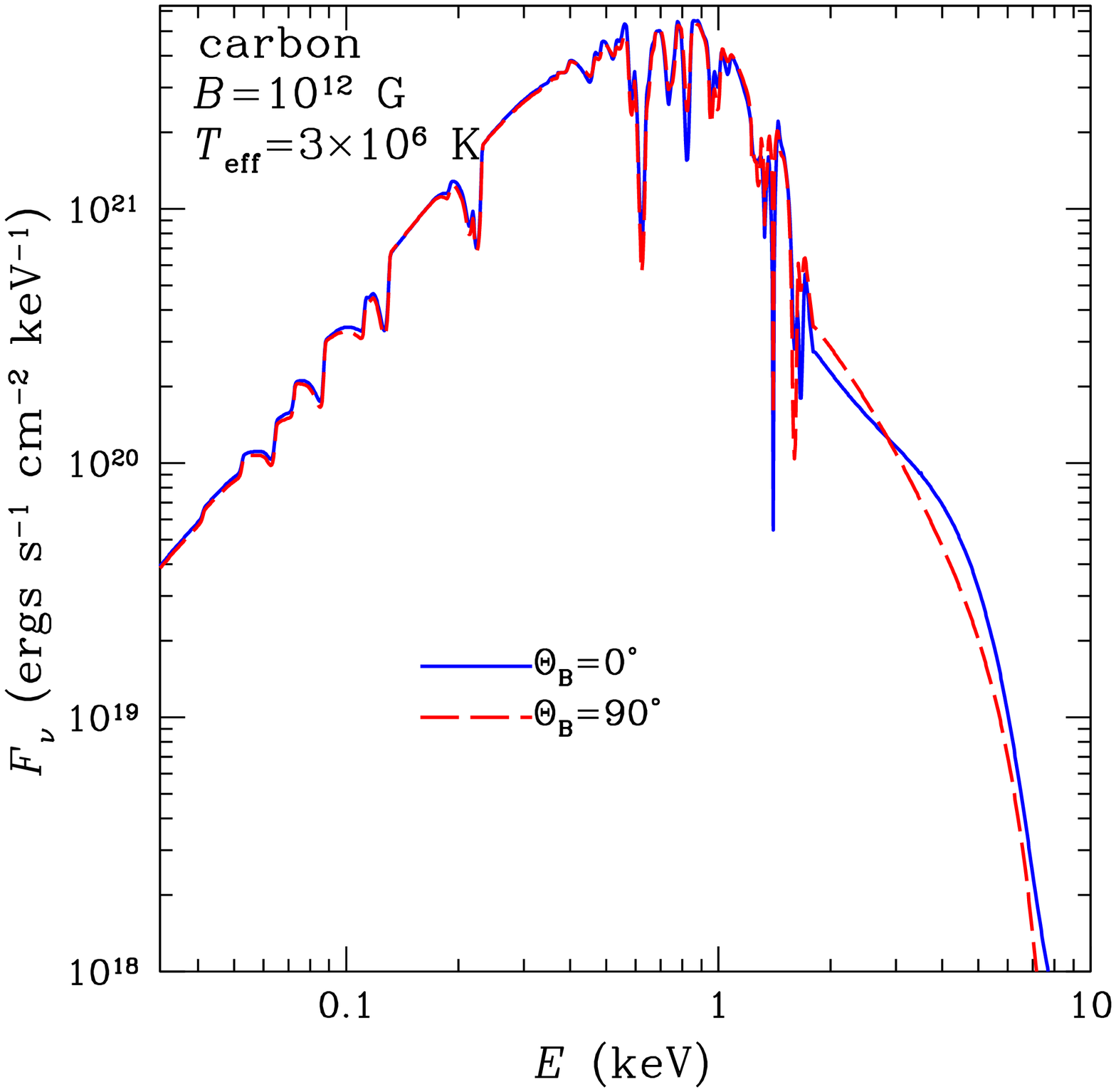}}
\caption{Temperature profile (top) and carbon atmosphere spectra (bottom)
with $\Theta_B=0^\circ$
(solid lines; magnetic field perpendicular to the atmosphere plane)
and $\Theta_B=90^\circ$
(dashed lines; magnetic field parallel to the atmosphere plane).
The other parameters are set to $B=10^{12}$~G and $\Teff=3\times10^6$~K.
\label{fig_spec_ang}}
\end{figure}

Here we illustrate the dependence of the atmosphere models on magnetic
field geometry ($\Theta_B$) and
surface gravity ($g$).  Figure~\ref{fig_spec_ang} shows
temperature profiles and carbon atmosphere spectra at two different
magnetic field angles ($\Theta_B$) with respect to the NS surface normal.
There is no remarkable difference
between the two magnetic field orientations, except at several strong lines and
the photo-ionization edges at high energy ($E\ga1.7$~keV for carbon at
$B=10^{12}$~G). The angular dependence shown here reflects the (rotational)
phase variation of atmosphere spectra, as the observer sees regions of the
NS surface with different magnetic field orientations
(see Section~\ref{sec_bfield}).

\begin{figure} 
 \resizebox{\hsize}{!}{ \includegraphics{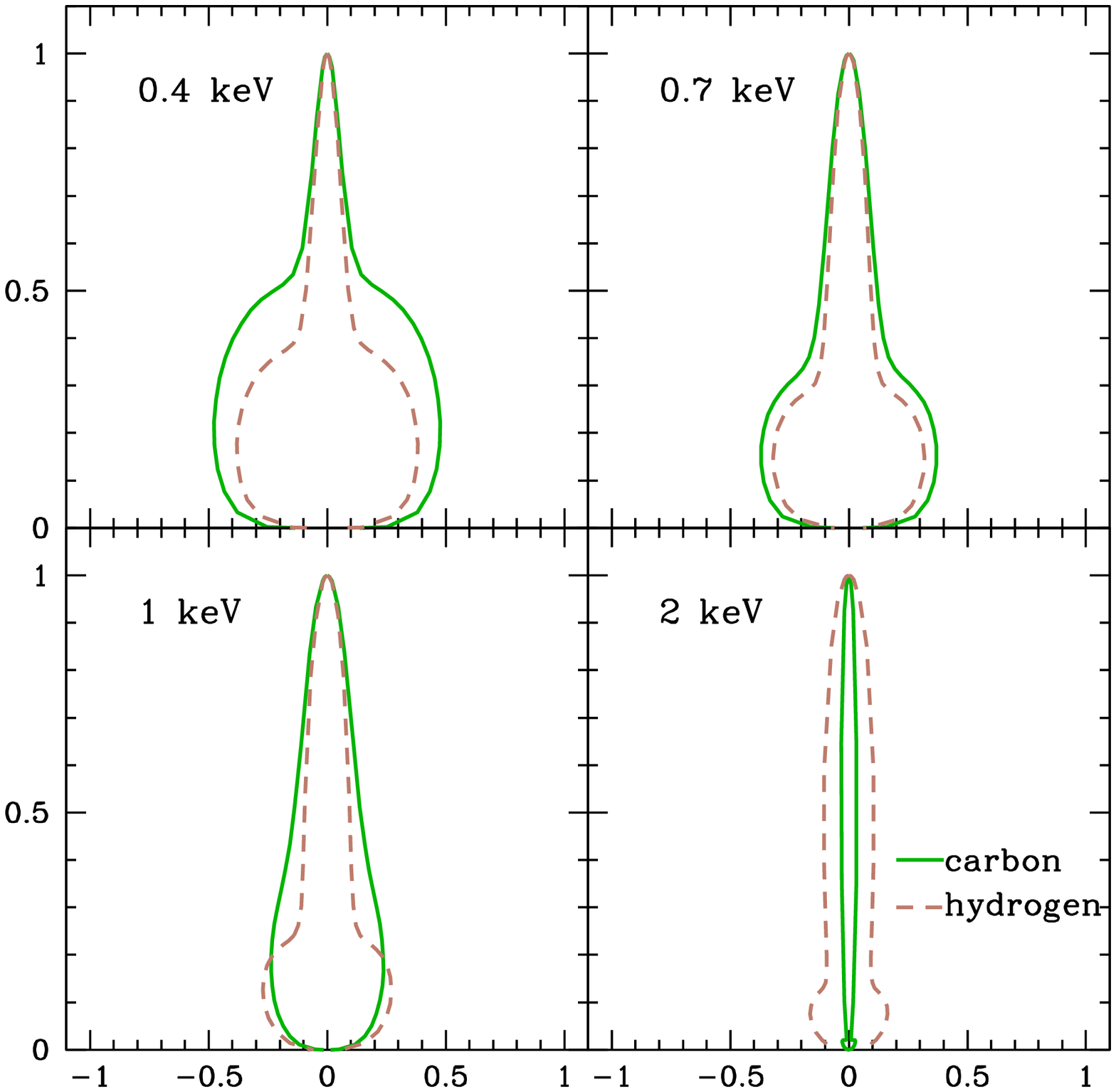}}   
\resizebox{\hsize}{!}{\includegraphics{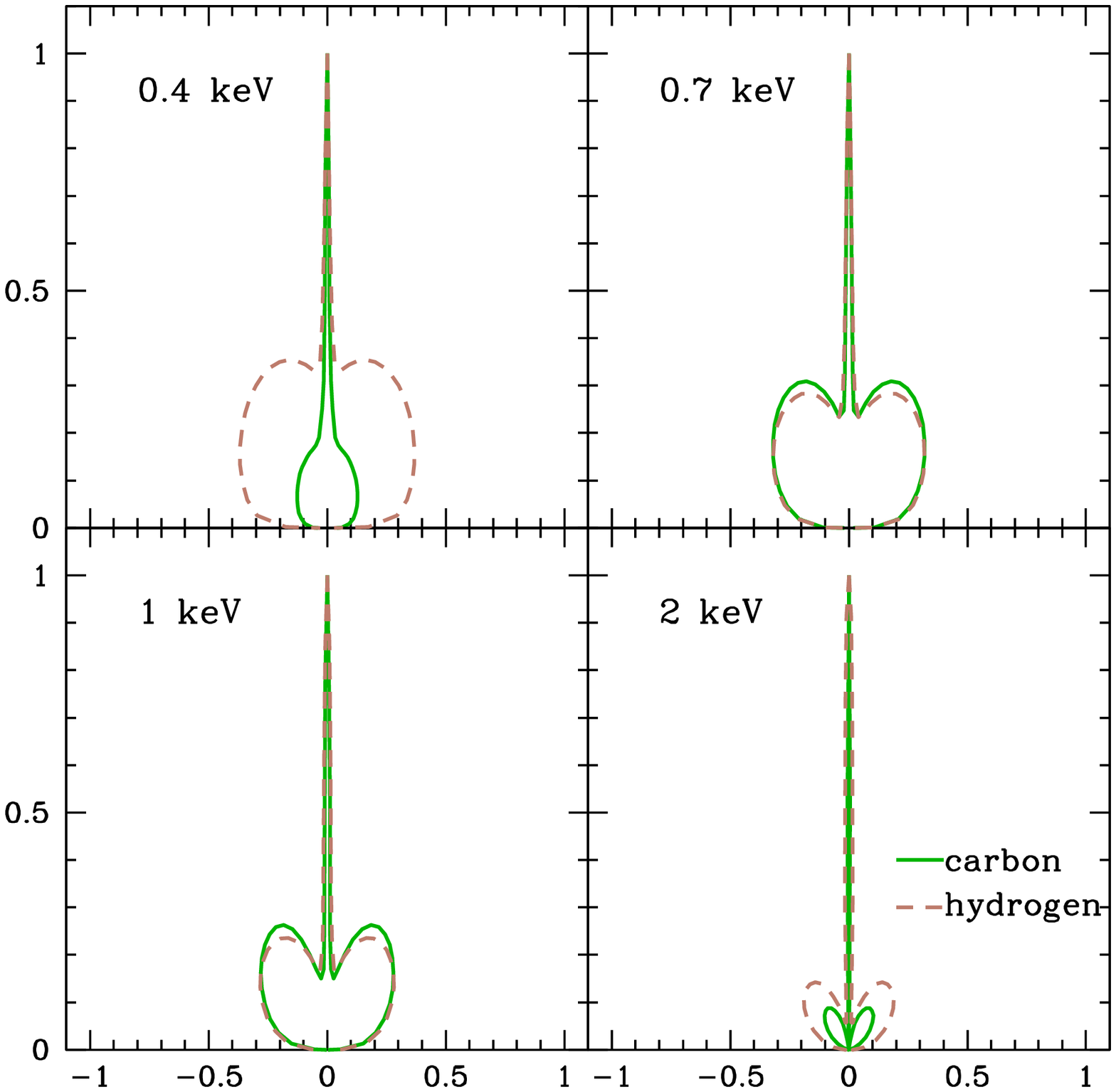}}
\caption{Polar diagrams of the specific intensity (scaled to the intensity
in the normal direction) from the carbon (solid lines) and hydrogen
(dashed lines) atmosphere models with $\Teff=3\times 10^6$~K and
$B=10^{12}$~G (top) and $B=10^{13}$~G (bottom).
\label{fig_beam}}
\end{figure}

Atmosphere emission can also be highly beamed in the presence
of a magnetic field \citep{shibanov92,pavlov94}.
Figure~\ref{fig_beam} shows the
energy-dependent beam pattern for the atmosphere models with
$\Teff=3\times 10^6$~K and $B=10^{12}$ and $10^{13}$~G.
Note that the emission is more beamed at higher magnetic fields. There
are some noticable differences between hydrogen and mid-Z element
atmospheres. Our results show that energy-resolved lightcurve analysis
can potentially be used as a diagnostic tool to determine the surface
composition. 

\begin{figure} 
\resizebox{\hsize}{!}{  \includegraphics{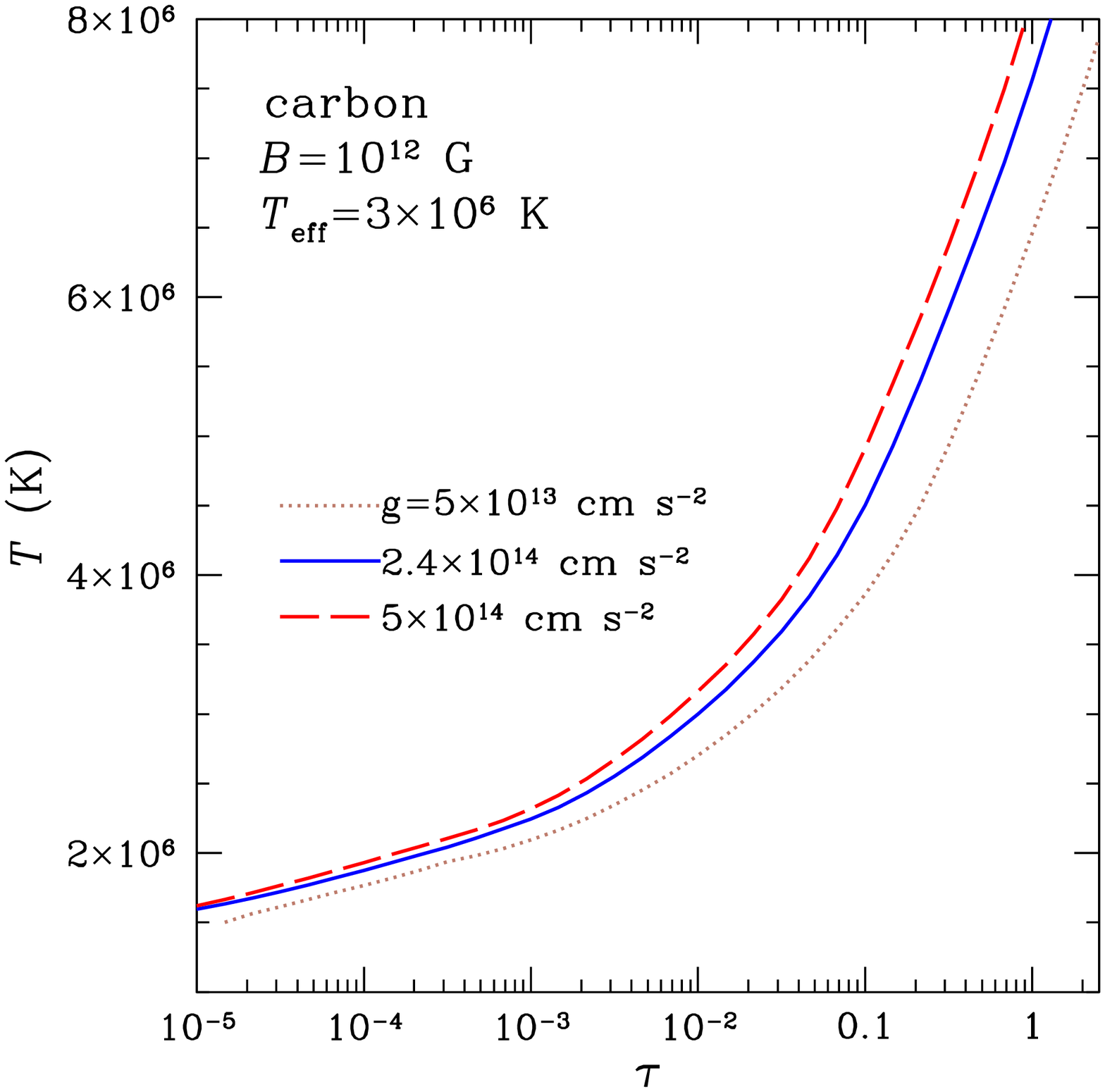}}  
\resizebox{\hsize}{!}{\includegraphics{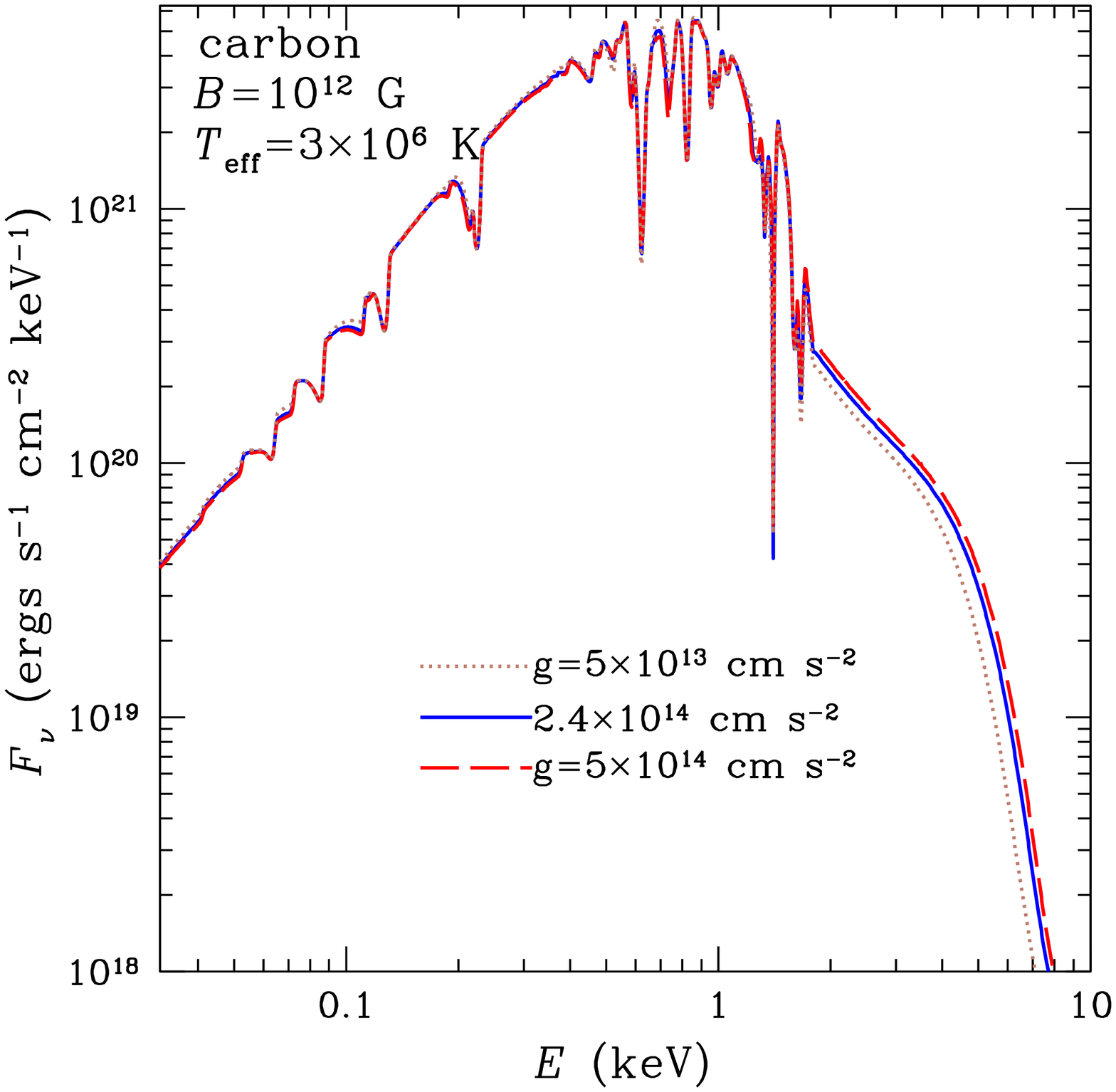}}
\caption{Temperature profiles (top) and carbon atmosphere spectra (bottom)
with $g=5\times10^{13}$ (dotted lines), $2.4\times 10^{14}$ (solid lines),
and $5\times10^{14}$~cm~s$^{-2}$ (dashed lines), corresponding
to ($M$,$R$) = ($0.5M_\odot$,$12$~km), ($1.4M_\odot$,$10$~km),
and ($2.2M_\odot$,$10$~km), respectively.
The other parameters are set to $B=10^{12}$~G, $\Teff=3\times10^6$~K,
and $\Theta_B=0^\circ$. \label{fig_spec_grav}}
\end{figure}

Figure~\ref{fig_spec_grav} shows carbon spectra at two rather extreme
values of $g$ ($=5\times10^{13}$ and
$5\times10^{14}$~cm~s$^{-2}$) and hence NS mass and radius
\citep[see, e.g.,][]{lattimer01}.
Our results show mid-Z element atmosphere spectra, like magnetic hydrogen
spectra \citep[see, e.g.,][]{pavlov95_1}, have a weak dependence on
gravitational acceleration. This is
contrary to the results of unmagnetized NS atmosphere spectra in which
several absorption features vary in their strengths and shapes strongly
with $g$ \citep{zavlin96}. The weak dependence on $g$ is partially due to the
fact that spectral features in mid-Z element atmospheres are primarily
broadened by motional Stark effects (which are independent of plasma
density) and pressure effects are small (since absorption lines are
formed at low density).

\section{X-ray spectroscopy and determination of neutron star parameters}
\label{sec_ns}

Given the results in the previous sections, it is useful to consider
what one can learn from X-ray spectroscopy of INSs, especially those
which show absorption features in their spectra. Four NS parameters are
important in determining thermal spectra; surface composition ($Z$),
magnetic field ($B$), effective temperature ($\Teff$), and
gravitational redshift ($\zg$). It is desirable to detect multiple
spectral features to break degeneracies between the parameters.
We discuss here the feasibility of measuring these NS parameters with
X-ray spectroscopy.


\subsection{Surface composition} 

Given the substantial differences in spectral energy distribution
(SED) and absorption features, it should be relatively easy to
distinguish between mid-Z element and hydrogen atmospheres. In
particular, a possible signature for the presence of a mid-Z element
atmosphere is the
detection of multiple strong absorption lines at high energy ($E\ga1$~keV).
Multi-wavelength SED analysis over optical and X-ray bands
will be useful since mid-Z element atmospheres are significantly
softer than hydrogen atmospheres \citep{pavlov96}.  Since the number
of absorption features and their energies are strongly dependent on
the surface composition, it is possible to distinguish between
carbon and oxygen atmospheres if multiple spectral features are
detected.

\subsection{Magnetic field} 
\label{sec_bfield}

\begin{figure}  
\resizebox{\hsize}{!}{\includegraphics{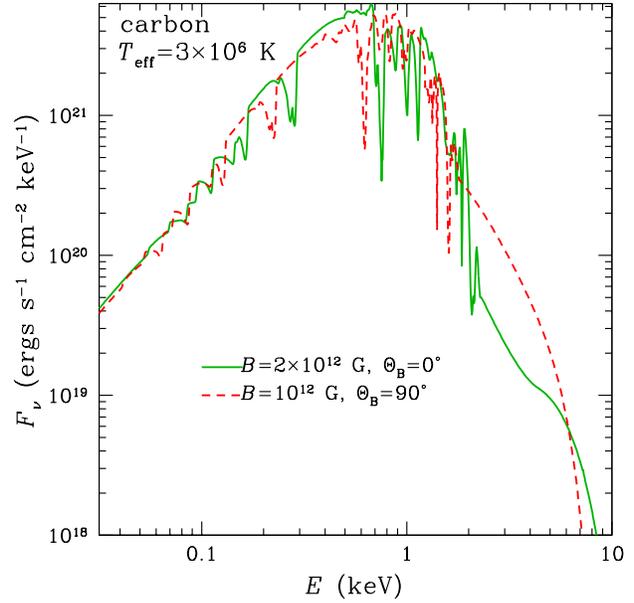}}
\caption{Carbon atmosphere spectra with $B=2\times10^{12}$~G and
  $\Theta_B=0^\circ$ (solid line) and $B=10^{12}$~G and
  $\Theta_B=90^\circ$ (dashed line). In both cases,
  $\Teff=3\times10^{6}$~K. The former and latter cases correspond to
  atmosphere spectra from the pole and equator in a dipole magnetic
  field configuration, respectively. \label{fig_mag_spec}}
\end{figure}

The absorption features shown in Figures~\ref{fig_spec_c12}--\ref{fig_spec_ne13}
are deeper and narrower than the observed spectral features from INSs
\citep{sanwal02, haberl03, haberl04, vankerkwijk04}.
This is partly due to the fact that our atmosphere models are constructed
with a single magnetic field strength $B$ and orientation $\Theta_B$, thus
reflecting emission from a local patch on the NS surface.
The absorption features in our model spectra will appear broadened if there
are magnetic field variations over the surface.
We expect to see dramatic changes in the spectra even with
a small magnetic field difference, since the ionization fraction and line
energies depend strongly on $B$. Figure~\ref{fig_mag_spec} shows
carbon atmosphere spectra at two magnetic field strengths and angles,
corresponding to the pole and equator for a magnetic dipole
configuration. Indeed, atmosphere spectra can look quite different merely
due to a factor of two difference in the magnetic field strength.
Thus thermal spectra of INSs can show significant rotational
phase-variation due to a magnetic field distribution, even when the
surface temperature is uniform.   

In order to construct synthetic atmosphere spectra, one must
integrate emission from patches of the NS surface with differing $B$,
$\Theta_B$, and $\Teff$.
Such synthetic spectra are necessarily model-dependent
\citep[see, e.g.,][]{zavlin95, zane01,ho04,zane06,ho06}, as the magnetic field and
temperature distributions over the surface are unknown.
Although magnetic field broadening is different for each absorption
line, atomic line energies vary approximately by
$\Delta E/E\simeq$ 10\% and 30\% for a factor of 1.5 and 2 variation
in magnetic field strength.  Electron or ion cyclotron lines may appear in the X-ray
band when $B$ is lower or higher than the values considered in this
paper. Some of the observed absorption features have been attributed
to cyclotron lines \citep{haberl03}. Since cyclotron line energy
varies linearly with $B$, we expect cyclotron lines to be broader and
have larger phase variation than atomic lines.  Therefore, line width
analysis can be a useful diagnostic tool to probe the magnetic field
distribution over the NS surface, given that the intrinsic widths of
absorption lines are narrow in mid-Z element atmospheres
\citep{mori06}. Magnetic field variation can smear out narrow absorption lines and
reduce the accuracy of determining other NS parameters such as the
gravitational redshift. Phase-resolved spectroscopy is essential to
disentangle magnetic field effects from the 
other parameters and to probe the magnetic field geometry.

\subsection{Effective temperature} 

The effective temperature must be measured to constrain the cooling
history of NSs \citep{page04,yakovlev04}.
Thermal spectra of
most INSs are fit with either blackbody or fully-ionized
hydrogen atmosphere models, each of which exhibits the softest and hardest
spectra, respectively. Therefore, errors in the measured effective
temperatures are
primarily due to uncertainty in the surface composition. 

Although mid-Z element atmospheres can have overall SEDs close to a blackbody, there
are several significant differences from blackbody spectra. In the 
Wien tail, mid-Z element atmosphere spectra are softer than a blackbody 
due to photo-absorption lines and edges and become harder than a
blackbody at higher energy. Therefore the effective temperature can be higher than the
blackbody temperature if high energy tails are not fully covered by X-ray
spectroscopy.  Alternatively, the high energy part of X-ray spectra may be
erroneously fit by an additional blackbody component
(see Section~\ref{sec_1207} for the case of 1E1207).  It is
important to cover a wide energy band and fit the data with model atmosphere
spectra, rather than fitting with one or two blackbody
components and simple Gaussian or Lorentzian absorption lines.

\subsection{Neutron star mass and radius} 

One of the ultimate goals in NS research is to uniquely determine the
mass and radius of NSs and constrain the nuclear
equation of state of the NS
core. In X-ray spectroscopy, gravitational redshift $\zg$ is one of
the NS parameters which can be measured once X-ray data
are fit with model atmosphere spectra.  The accuracy of a gravitational
redshift measurement is limited by magnetic field surface variations since
line energies change with both $B$ and $\zg$. To increase
accuracy, phase-resolved spectroscopy is essential because $B$ varies with
rotation phase but $\zg$ does not. Grating spectroscopy will measure
gravitational redshift with high accuracy if magnetic field variation
is not so severe and photon statistics is good. In the future,
high-resolution spectroscopy missions, such as 
{\it Constellation-X}, will be able to measure the magnetic field and
gravitational redshift with unprecedented accuracy.

It is possible to measure the surface gravity ($g$) from
line widths if the lines are primarily broadened by pressure effects
\citep{paerels97}.  However, motional Stark effects produce line
broadening far larger than pressure effects; this is partly because lines
are formed in low density layers.  Also, our study indicates that it
is difficult to measure surface gravity from overall
spectral fitting since mid-Z element spectra have a very weak dependence
on $g$ (see Figure~\ref{fig_spec_grav}). Therefore, the methods for
constraining $g$ by using absorption features is applicable only to
weakly-magnetized NSs ($B\la10^9$ G) since their spectra are nearly
free from the magnetic field effects discussed in this paper
\citep[see also][]{zavlin96}.
On the other hand, it may be possible to constrain $g$ in heavier element 
atmospheres (e.g., Fe) since motional Stark effects are further
reduced while ion-coupling effects are enhanced. We plan to
investigate heavy element atmospheres composed of silicon and iron in
future work. 

Radius measurements are feasible by combining X-ray
spectroscopy (which measures flux) and distance measurements. Distances
to several nearby INSs have been measured with relatively high accuracy by
parallax measurements \citep{walter02, kaplan02,
vankerkwijk06}. However, it should be kept in mind that only a
fraction of the surface (such as hot spots) may be visible due to
temperature anisotropy caused by strong magnetic fields
\citep{greenstein83, geppert04, geppert06}. Independent lightcurve
analysis will be helpful to determine the size of the emission region.


\section{Comparison with X-ray data of isolated neutron stars} 
\label{sec_obs}

In this section, we discuss INS thermal spectra in the context of mid-Z
element atmospheres. We consider the INS 1E1207.4$-$5209 and the class of
radio-quiet INSs; these all show spectral features
in their X-ray spectra, except RX~J1856.5$-$3754. Table~\ref{tab_obs}
summarizes their observational properties based on recent \chandra\ and \xmm\
observations. 

\begin{table*}
\centering 
\caption{Summary of X-ray observations of 
selected INSs \label{tab_obs}} 
\begin{tabular}{@{}lccccc@{}}
\hline 
Object & $kT_{BB}$ [eV] & $B_d$ [$10^{13}$ G] & $E_{line}$ [keV] & $\sigma$ [eV] & EW
  [eV] \\ 
\hline
1E1207.4$-$5209$^{({\rm a})}$ & 260 & ?  & 0.7, 1.4 & 130, 76 & 100, 60
\\ 
RBS 1774 & 102 & $< 24$ & 0.75 & 27 & 27 \\ 
RX J1308.6$+$2127 & 103 & 3.4 & 0.23, 0.46 & 150, 260 & 200, 180 \\
RX J1605.3$+$3249 & 96 & ? & 0.40, 0.59, 0.78 & 87$^{({\rm f})}$ & 96, 76, 67 \\
RX J0806.4$-$4123$^{({\rm b})}$ & 96 & $<14$ & 0.43 & 70$^{({\rm f})}$ & 33 \\ 
RX J0720.4$-$3125$^{({\rm c})}$ & 85--95 & 2.4 & 0.28 & 90 & 0--70 \\ 
RX J1856.5$-$3754 & 62 & $\sim1$$^{({\rm d})}$ & --- & --- & --- \\ 
RX J0420.0$-$5022$^{({\rm e})}$  & 45 & $<18$ & 0.33 & 70$^{({\rm f})}$ & 45\\
\hline 
\end{tabular} 
\begin{quote} 
 {\scshape Notes:}\\
We list blackbody temperature ($T_{BB}$), dipole
magnetic field strength ($B_d$), line energy, line width, and equivalent
width of the observed absorption features. All the temperatures were
measured by assuming a single blackbody component. The observed
absorption lines were fit with a Gaussian line profile and $\sigma$ refers to
the Gaussian line width defined in equation (1) of \citet{mori05}. The line parameters are
subject to the choice of continuum models \citep{mori05} and the number of fitted lines 
\citep{haberl06_2}. See \citet{haberl06_2} for a complete reference list
of the observations.\\ 
(a) The
source shows an excess above a blackbody at $E\ga2$ keV; an additional
continuum component has been used to fit this excess \citep{mori05}. \\ 
(b) An additional
absorption feature may be present at higher energy, and spectral
fitting with two lines yields 
$E_{line}=0.31$ and 0.61~keV \citep{haberl06_2}.\\ 
(c) Over the last six years, X-ray
spectra of RXJ0720 have become harder, and the line strength  has
increased \citep{devries04, vink04, haberl06}.\\ 
(d) Dipole magnetic field strength was estimated from
the H$\alpha$ nebula in the vicinity of RX~J1856.5$-$3754
\citep{vankerkwijk01_2, kaplan02}. \\ 
(e) Based on the 
20 ksec \xmm\ EPIC-PN data, \citet{haberl04_2} indicated an absorption
feature at $E=0.33$ keV.  However, its significance is not 
definitive yet \citep{haberl06_2}, and our independent analysis of the
\xmm\ data has not confirmed a significant spectral feature. \\ 
(f) The line width was fixed while fitting the spectra. 
\end{quote} 
\end{table*}

\subsection{1E1207.4$-$5209}
 \label{sec_1207}

1E1207 is a hot isolated NS\footnote{Recent timing analysis suggests
  that 1E1207 is in a binary system \citep{woods06}. However, it is
  unlikely that 1E1207 is an accreting NS.} with an age~$\sim7\times10^3$~y.
Remarkably, it shows two broad absorption features at $\sim0.7$
  and $\sim1.4$~keV and an excess above blackbody emission at $\ga2$~keV.
A second blackbody component with a small emitting area ($R \la 1$~km)
and high temperature ($T\ga 3\times10^6$~K) has been used to fit the high
energy excess \citep{mereghetti02, bignami03, deluca04,
  mori05}. However, it is puzzling that the pulsed fraction is not
  particularly large at high energies \citep{deluca04}, while
  small hot spots can strongly modulate X-ray lightcurves.
Also, the temperature can be quite different depending on the
  number of continuum components used to fit the data, e.g.,
$kT_{BB}$=260~eV when using a single blackbody and
$kT_{BB}$=150~eV when using a double blackbody
\citep[the second blackbody in the latter case has $\sim$300~eV;][]{mori05}.

Proposed models involving
  cyclotron lines or atomic lines from a light element (H, He)
  atmosphere \citep{sanwal02, bignami03, turbiner04_1} seem
  unlikely since line strengths are significantly weakened
  by vacuum resonance effects
\citep[and references therein]{ho04,vanadelsberg06} or the
  ionization states responsible for the observed lines are not
  abundant \citep{mori06}. Alternatively, \citet{hailey02} and
  \citet{mori06} proposed a mid-Z element atmosphere for 1E1207.

\begin{figure} 
 \resizebox{\hsize}{!}{ \includegraphics{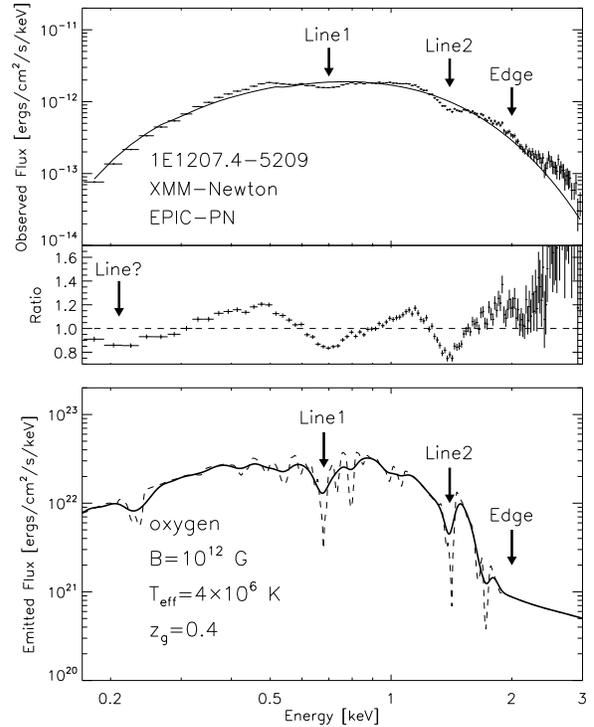}}
\caption{Unfolded \xmm\ EPIC-PN spectrum of 1E1207.4-5209 (top) and spectra of oxygen atmosphere with
  $B=10^{12}$~G, $\Teff=4\times10^6$~K and $\zg=0.4$ (bottom). In the
  top panel, we overlaid a blackbody spectrum to illustrate the
  absorption features; we also show the ratio of the data and blackbody.
  There is potentially another absorption feature at
  $E\la0.3$ keV (indicated by an arrow in the ratio plot).  
In the bottom panel, the solid line is the model atmosphere spectrum convolved with the
\xmm\ EPIC-PN energy resolution, while the dashed line is the raw spectrum.
\label{fig_1207}}
\end{figure}

Figure \ref{fig_1207} shows the \xmm/EPIC-PN spectrum of 1E1207 and the oxygen atmosphere spectrum
with $B=10^{12}$~G, $\Teff=4\times10^6$~K and $\zg=0.4$; the latter
is convolved with the \xmm\ EPIC energy resolution. 
Although many of the narrow absorption lines are blended
or smeared out, some substructure is still seen. Remarkably, the oxygen
atmosphere spectrum in Figure~\ref{fig_1207} exhibits the three
spectral features seen in the 1E1207 spectrum (i.e., two broad absorption
features at $\sim 0.7$ and 1.4~keV and spectral hardening above $\sim2$
keV). See Table~1 in \citet{mori06}
for identification of the bound-bound transitions responsible for
the 0.7 and 1.4~keV features and their unredshifted line
energies. In our convolved model spectrum, the 0.7~keV feature shows a
more complex structure than the 1.4~keV feature because there are more
blended absorption lines at $\sim 0.7$~keV. Indeed, some analyses of the
\xmm/EPIC data of 1E1207 indicate the presence of substruture at
$\sim0.7$~keV \citep{mereghetti02, mori06}. The 
edge-like feature in the model spectrum at $\sim2$~keV (similar to the
excess above a blackbody in the 1E1207 spectrum) is 
due to bound-free transitions from the innermost electron
in the $(m\nu)=(00)$ state of various oxygen ions. In the 1E1207
spectra, there is an indication of an additional absorption feature
at $E\la0.3$~keV,
where the oxygen atmosphere spectrum also shows an absorption
line. Given the less reliable instrumental calibration and stronger
effects of interstellar absorption at low energies, we plan to more
carefully investigate the significance of this potential absorption feature in
future work.  

Note that the model atmosphere spectrum shown here is not completely
realistic because we have not
taken into account the distribution of magnetic field and temperature
over the surface (see Sections~\ref{sec_mag_grav} and \ref{sec_bfield}).
More detailed analysis based on spectral fitting by varying NS atmosphere
parameters and viewing geometries is in progress.
Given the good photon statistics with the 260 ksec \xmm\ data
\citep{bignami03, mori05} and the three (or possibly four) spectral features to fit, we are optimistic
about pinning down the surface composition, magnetic field, temperature,
and gravitational redshift of 1E1207 with the existing
CCD spectral data.

\subsection{Radio-quiet neutron stars}

Seven nearby INS are characterized as radio-quiet NSs (RQNSs) since they
exhibit similar spectral and timing properties
\citep{vankerkwijk06}. At least five of them show a single or multiple
absorption features at $E\simeq0.2-0.7$~keV \citep{haberl03,
vankerkwijk04, haberl04, zane05, schwope05, haberl06_2}, while the brightest
INS RX~J1856.5$-$3754 shows a nearly perfect blackbody spectrum without any
spectral features \citep{burwitz02}. Some of the observed absorption
features have been interpreted as proton cyclotron lines or
hydrogen/helium atomic lines at $B=10^{13}$--$10^{14}$~G
\citep{vankerkwijk06}.  Timing analyses yield similar dipole magnetic
field strengths \citep{kaplan05_1, kaplan05_2}.

\begin{figure} 
 \resizebox{\hsize}{!}{ \includegraphics{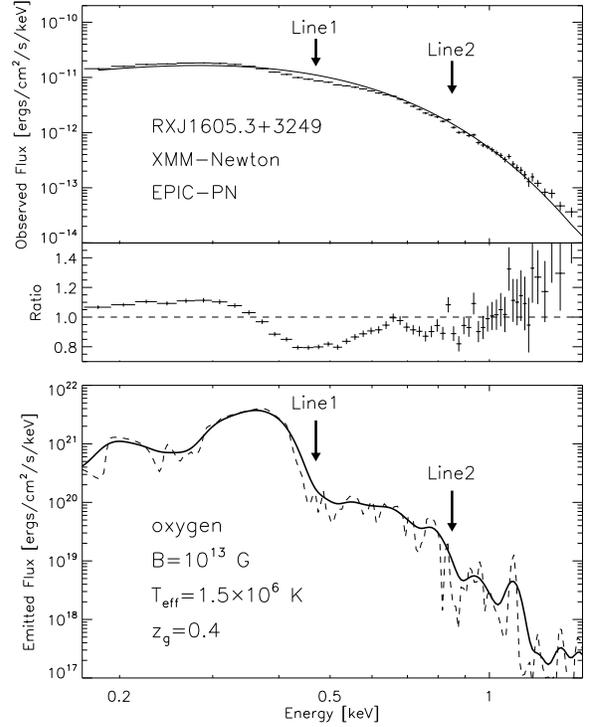}}
\caption{Unfolded \xmm\ EPIC-PN spectrum of RXJ1605.3+3249 (top) and spectra of oxygen atmosphere with
  $B=10^{13}$~G, $\Teff=1.5\times10^6$~K and $\zg=0.4$ (bottom). In the top
  panel, we overlaid a blackbody spectrum to illustrate the absorption
  features; we also show the ratio of the data and blackbody.
In the bottom panel, the solid line is the model atmosphere spectrum convolved with the
\xmm\ EPIC-PN energy resolution, while the dashed line is the raw spectrum.
 \label{fig_rqns}}
\end{figure}

Figure~\ref{fig_rqns} shows the \xmm/EPIC-PN spectrum of RX~J1605.3$+$3249
and the oxygen atmosphere spectrum with
$B=10^{13}$~G, $\Teff=1.5\times10^6$~K and $\zg=0.4$; the latter is convolved
with \xmm\ EPIC energy resolution. 
We use the spectrum of RX~J1605.3$+$3249 as an illustration of the
X-ray spectra of several other RQNSs with absorption features at
$E\sim0.3$--0.4\ keV (see Table~\ref{tab_obs}).  In addition to the
feature in the model spectrum at (redshifted) 0.4~keV, there are
weaker absorption features at higher energies; these may be present in
the spectrum of RX~J1605.3$+$3249 or RX~J1308.6$+$2127 
\citep{haberl06_2}. If all RQNSs have similar magnetic field strengths
(e.g., $B\simeq\mbox{few}\times10^{13}$~G, as suggested by timing
analyses), we may expect RQNSs with higher temperatures to have more distinct
absorption features at high energy (see, e.g., Figure~\ref{fig_spec_ne13}).
This appears to agree with the
observed properties listed in Table~\ref{tab_obs}.  Further X-ray
observations are needed to confirm the significance of multiple
absorption lines.

On the other hand, it is puzzling as to why the spectrum of
RX~J1856.5$-$3754 can be fit by a featureless blackbody spectrum.
Since RX~J1856.5$-$3754 has a lower temperature than most of the other RQNSs,
it is possible that molecular chains are formed on the surface
\citep{medin06_1, medin06_2}, or RX~J1856.5$-$3754 has the condensed surface
\citep{turolla04, vanadelsberg05, perez05}. Recent theoretical studies
indicate that the condensed surfaces of magnetized NSs exhibit blackbody-like
spectra in the X-ray
band \citep{turolla04, vanadelsberg05, perez05}.  At the observed
temperature of RX~J1856.5$-$3754 ($T_{BB}\sim7\times10^5$~K), it is estimated
that a carbon atmosphere may become condensed when $B\ga3\times10^{13}$~G
\citep{lai01, medin06_2}.  In this context, we expect that the other
RQNS (RX~J0420.0$-$5022) with a low temperature ($T_{BB}\sim5\times10^5$~K)
has a condensed surface; RX~J0420$-$5022 should thus show a featureless
blackbody-like spectrum \footnote{See caption (e) of table 1 for the
  current status of the putative line feature in the RX~J0420.0$-$5022 spectrum.}, unless it has a surface magnetic field lower than
$10^{13}$~G. Indeed, the sharp temperature dependence of the molecular
abundance was shown in the case of a helium atmosphere
\citep{mori06_3}. 

Optical flux excess from Rayleigh-Jeans tails of X-ray blackbody
spectra is another common signature of RQNSs
\citep{burwitz02}. Fully-ionized hydrogen atmosphere models overpredict
the optical fluxes
by 1--2 orders of magnitude \citep{pavlov96, pons02}. Unmagnetized heavy
element atmosphere models can account for the optical excess \citep{pons02},
but they also exhibit numerous absorption features, which significantly
deviate from the featureless X-ray spectra of RX~J1856.5$-$3754
\citep{zavlin96,
rajagopal96, gansicke02}.  Therefore, it requires either a thin hydrogen
atmosphere \citep{ho06} or an anisotropic temperature distribution over
the surface \citep{pons02, braje02, perez06} to produce both the
featureless X-ray spectrum and the optical excess.  Since magnetized
mid-Z element atmospheres have SEDs close to blackbody, they may
account for the optical excess. Unlike unmagnetized atmospheres, magnetic
field variations can smear out absorption
features in magnetized mid-Z atmosphere spectra and produce
featureless X-ray spectra.

\section{Summary}
 \label{sec_summary}

We have constructed equation of state and opacity tables for mid-Z
elements using the most up-to-date methods available and improving upon
earlier calculations by including more physical effects.
Using these tables, we built magnetic neutron star atmosphere models for carbon,
oxygen, and neon.  Mid-Z element atmosphere spectra show some dependence
on magnetic field orientation (relative to the surface) and a weak
dependence on surface gravity.  There are many characteristics that
distinguish these spectra from hydrogen atmosphere spectra, e.g.,
a multitude of lines, different line energy and strength dependences
on the magnetic field, and possible hardening at high energies.

There are still several issues that need to be addressed in order to improve
our models.  These include accounting for molecules, the quantum
electrodynamical effect of vacuum resonance, and non-LTE effects.
Molecules and vacuum resonance may be important for atmosphere models with
magnetic fields higher than those considered in this paper, i.e.,
$B\gg 10^{13}$~G \citep{ho06_2}.
We also plan to study atmosphere models with different
elements, such as helium, silicon and iron, in order to cover a larger
NS atmosphere parameter space.  We will construct synthetic atmosphere
model spectra by taking into account magnetic field and temperature distributions
on the surface. 

We presented qualitative comparisons between our model spectra and the
observations of several isolated neutron stars.  The results are
promising; 
the model atmosphere spectra closely resemble the X-ray spectra of these
objects.  The similarities strongly motivate more detailed studies,
including phase-resolved analysis, since line identification and a better
temperature estimation can provide invaluable information on neutron stars.

\section*{acknowledgements} 

KM thanks Dong Lai for the hospitality at Cornell University where
this work was initiated. We are indebted to Dong Lai and Alexander
Potekhin for their continuous support and many useful discussions. We
also appreciate valuable comments from Marten van Kerkwijk, George
Pavlov and Chris Thompson.  We thank the anonymous referee for comments that
  helped to improve the paper. 
WH is grateful for the use of the computer facilities at the Kavli
Institute for Particle Astrophysics and Cosmology.
WH is partially supported by NASA through Hubble Fellowship grant
HF-01161.01-A awarded by the Space Telescope Science Institute, which
is operated by the Association of Universities for Research in Astronomy,
Inc., for NASA, under contract NAS~5-26555.


\label{lastpage}

\end{document}